\documentclass[final,3p,times,twocolumn]{elsarticle}

\usepackage{epsfig,amsmath,amsbsy,amssymb,bbm}

\usepackage{color,colordvi}
\definecolor{Red}{rgb}{0.9,0.0,0.1}
\definecolor{Blue}{rgb}{0.1,0.0,0.9}
\definecolor{Black}{rgb}{0.0,0.0,0.0}

\renewcommand{\vec}[1]{\boldsymbol{#1}}
\newcommand{\op}[1]{\mathsf{#1}}

\journal{J. Colloid Interface Sci.}

\begin{document}

\begin{frontmatter}

\title{Pendant Capsule Elastometry}

\author[label1]{Jonas Hegemann}
\author[label1]{Sebastian Knoche}
\author[label2]{Simon Egger}
\author[label2]{Maureen Kott}
\author[label2]{Sarah Demand}
\author[label2]{Anja Unverfehrt}
\author[label2]{Heinz Rehage}
\author[label1]{Jan Kierfeld}

\address[label1]{Physics Department, TU Dortmund University, 44221 Dortmund, Germany}
\address[label2]{Department of Chemistry, TU Dortmund University, 44221 Dortmund, Germany}

%%%%%%%%%%%%%%%%%%%%%%%%

\begin{abstract}
We provide a C/C++ software for 
the shape analysis of deflated elastic capsules in a
pendant capsule geometry, which is based on an  elastic description of
the capsule material as a quasi two-dimensional elastic membrane using shell
theory. 
Pendant capsule elastometry provides a new in-situ and non-contact 
method
for interfacial rheology of elastic capsules that goes beyond 
determination of the  Gibbs- or dilational modulus from 
area-dependent  measurements of the surface tension using 
pendant drop tensiometry,
which can only give a rough estimate of the elastic capsule properties as they
are based on a purely liquid interface model.
Given an elastic model of the capsule membrane, pendant capsule elastometry 
determines optimal elastic moduli by fitting numerically generated 
axisymmetric shapes optimally to an experimental image. 
For each digitized image of a  deflated capsule elastic moduli 
can be determined, if another image of its undeformed 
reference shape is provided.
Within this paper, we focus on nonlinear Hookean elasticity because of 
its low computational cost its wide applicability, 
but also discuss and implement alternative constitutive laws.
For Hookean elasticity, Young's
surface modulus (or, alternatively, area compression modulus) 
and  Poisson's ratio are determined; for Mooney-Rivlin 
elasticity, the Rivlin modulus and a dimensionless shape parameter 
are determined; for neo-Hookean
elasticity, only the Rivlin modulus is determined, using a fixed
dimensionless shape parameter.
Comparing results for different models we find that 
 nonlinear Hookean elasticity is adequate for most capsules. 
If series of images are available,
these moduli can be evaluated as a function of the capsule volume
to analyze hysteresis or aging effects depending on the deformation history
or to detect viscoelastic effects for different volume change rates. 
An additional wrinkling wavelength measurement allows the user 
to determine the  bending modulus, from which 
the layer thickness can be derived. 
We verify the method by analyzing several materials, compare the results to
available rheological measurements, and review several applications.
We make the software available under the GPL license at 
github.com/jhegemann/opencapsule.
\end{abstract}


\end{frontmatter}

%% \linenumbers

%%%%%%%%%%%%%%
\section{Introduction}

Elastic capsules that consist of a solid thin shell enclosing a liquid volume 
  can be produced artificially by a variety of chemical processes,
such as interfacial crosslinking or polymerization
  \cite{Rehage2002}.
Moreover, solid-like interfaces can form by interfacial adsorption and 
self-assembly of surface active micro- or nano-particles  such as
colloidal particles in colloidosomes \cite{Dinsmore2002},
petroleum \cite{Freer2003}, various proteins at interfaces 
\cite{Noskov2011,Noskov2014}, for example 
hydrophobins at water-air interfaces \cite{Aumaitre2013}.
Eventually, solid-like shells can likewise be formed using layer-by-layer
assembly by employing electrostatic interactions 
of polyelectrolytes \cite{Donath1998,Guzman2009,Cramer2017}.
Elastic capsules have 
many applications for transport and delivery of the enclosed liquid 
in pharmaceutical, cosmetic or chemical industry \cite{Neubauer2013}.
Likewise, they  serve as biological model systems for
red blood cells or the cell cortex. 
For all applications, a characterization of the mechanical 
properties of the capsule shell, i.e., its elastic moduli, 
is necessary \cite{Vinogradova2006,Neubauer2013}.

Encapsulation applications employ closed microcapsules,
but often capsules can likewise be produced in a pendant or hanging capsule 
geometry, where the capsule is not closed and the capsule edge 
is attached to a capillary 
\cite{Ferri2010,Carvajal2011,Ferri2012,erni2012interfacial,Alexandrov2012a,Aumaitre2013,danov2015capillary,Salmon2016,Cramer2017}. 
Such capsules can be produced by self-assembly onto a droplet hanging 
from a capillary or onto an air bubble rising from a capillary,
or by interfacial crosslinking at the interface of a pendant droplet
\cite{knoche2013elastometry}. 
An advantage of this pendant capsule geometry is that 
volume reduction or pressure application can easily be realized 
by fluid suction through the capillary and it, thus,
offers a simple way of micromanipulation for mechanical characterization.

The related pendant droplet tensiometry is a standard tool 
to determine the surface tension of a
liquid interface using the Laplace-Young 
equation to model the droplet shape
\cite{Andreas1937,stauffer1965measurement,cabezas2006determination,berry2015measurement},
which is commercially available. 
The same Laplace-Young analysis has  frequently 
 been applied to pendant elastic capsules with different 
shell materials or droplets coated with  solid-like layers of adsorbed particles
\cite{Freer2003,Russev2008,Guzman2009,Ferri2010,Miller2010,Kovalchuk2010,Noskov2011,erni2012interfacial,Alexandrov2012a,danov2015capillary}
resulting in the determination of an ``effective surface tension'' $\gamma$ 
 describing  the solid shell interface of surface area $A$.
Changing the surface area $A$ in deflation experiments,  
the so-called Gibbs- or dilational modulus
$E_\mathrm{Gibbs} = \mathrm{d}\gamma/\mathrm{d}\ln A$ can be calculated. 
Pendant drop tensiometry can also be applied to droplets or capsules
with a viscoelastic interface by employing oscillating droplets
\cite{Russev2008,Ferri2010,Miller2010,Kovalchuk2010,Noskov2011};
then a complex dilational modulus can be 
obtained, which includes a real  elastic and 
an imaginary loss part.
The elastic dilational modulus  is equal to the area
 compression modulus $K_\mathrm{2D}$ for a fluid interface 
or for a two-dimensional solid interface
 in a planar Langmuir-Blodgett trough geometry. 
Application of the same 
concept to pendant elastic capsules gives 
misleading results because of inhomogeneous and anisotropic
elastic stresses in the capsule geometry and  
the existence of a curved undeformed reference shape of the capsule 
\cite{Ferri2010,Carvajal2011,Ferri2012,knoche2013elastometry,Cramer2017}. 
In Ref.\ \cite{knoche2013elastometry}, an elastic model 
based on shell theory 
has been developed which is capable of describing
capsule shapes in a deflation experiment 
more realistically. 
Similar  elastic models have been formulated in
  Refs.\ \cite{Ferri2010,Carvajal2011,Ferri2012,Cramer2017}.  
In Ref.\ \cite{knoche2013elastometry} this approach has been 
extended to the  pendant capsule elastometry method, where 
 the elastic model is 
 used  to determine two elastic constants, the surface Young modulus
$Y_\mathrm{2D}$ and Poisson's ratio $\nu_\mathrm{2D}$, by 
optimally  fitting calculated  shapes to  experimental images.
Pendant capsule elastometry has already been applied to OTS-capsules and 
hydrophobin-coated bubbles \cite{knoche2013elastometry} but also 
to bacterial films at interfaces \cite{Vaccari2015}.

Here, we want to present and make publicly available 
a much more efficient implementation of the pendant 
capsule elastometry method as a C/C++ software with a 
high degree of numerical efficiency and automation. 
In contrast to Ref.\ \cite{knoche2013elastometry}, where 
elastic constants were optimized on a  grid  in parameter space
to optimally match the experimental shape profile, 
we optimize elastic constants  in continuous parameter space,
which improves both performance and accuracy.
Moreover, we go beyond Ref.\ \cite{knoche2013elastometry} and 
generalize the shape analysis method to other constitutive laws.
In particular we investigate the behavior of the shape analysis
method in combination with Mooney-Rivlin or neo-Hookean elasticity
models, which are commonly used for inextensible polymeric materials.

These significant improvements turn the analysis into a strong
tool to investigate different materials in a short time and
on a large scale. 
We demonstrate these capabilities by analyzing a variety 
of deformation experiments for different materials.
In pendant capsule elastometry  Young's
modulus and  Poisson's ratio (or the Rivlin modulus
and the dimensionless shape parameter) of the two-dimensional 
capsule shell material are obtained from an
 analysis  of a digitized image of the 
deflated capsule shape and a second 
image of its undeformed reference shape. 
If the capsule wrinkles upon deflation,
an additional wrinkling wavelength measurement allows us 
to determine the  bending modulus, from which  the  layer
thickness can be derived if the shell material is a thin layer
of a three-dimensional isotropic elastic material.

%%%%%%%%%%%%%%%%%%%%%%%%%%
\section{Available experimental methods}

Several interfacial rheology methods exist, which allow the determination 
of the elastic properties of the capsule shell material.  
We review four different rheological methods,
which we will use as references for the pendant capsule method described in
this paper. Typical experimental methods are (i) surface shear-rheometry 
\cite{Erni2003}, (ii)
Langmuir-Blodgett trough, (iii) shear flow rheoscope (flow cell)
\cite{koleva2012deformation}, and (iv) spinning drop apparatus
\cite{pieper1998deformation}. 
Methods (i) and (ii) work with {\it planar} membranes of the shell 
material, whereas methods (iii) and (iv) directly work in the curved 
capsule geometry, like pendant capsule elastometry does. 
Apart from these four methods there are other contact 
techniques such as 
probing capsules with  AFM tips, micromanipulators, or optical tweezers
(see Ref.\ \cite{Neubauer2013} for a review).  
Pendant capsule elastometry is a non-contact technique and,
in comparison with methods (iii) and (iv), it does not require 
fluid motion in the surrounding fluid. 
We focus here on elastic capsule shell materials. For viscoelastic 
materials there are other interfacial rheology 
 methods available \cite{Miller2010}, 
such as double wall ring rheometry 
\cite{Vandebril2010} or magnetic rod rheometry 
\cite{Reynaert2008}.

In shear-rheometry, a transducer (thin disk or ring) is placed in a circular
vessel at a planar liquid-liquid or
air-liquid interface; between transducer and container wall a membrane 
with the shell material is prepared, such that
membrane deformations can be applied in circumferential direction.  While
oscillating at a certain frequency, the mechanical response is measured,
which gives the interfacial storage modulus $\mu'$ 
and the loss modulus $\mu''$. 
From $\mu'$ one determines the surface Young modulus $Y_\mathrm{2D} =
2(1+\nu_\mathrm{2D})\mu'$ provided that the Poisson ratio $\nu_\mathrm{2D}$ 
is known.  
%The method gives
%reproducable results and allows us to differentiate between viscoelastic and
%elastic behavior.  

In a Langmuir-Blodgett trough,
 a membrane made from the shell material 
is prepared in a rectangular vessel at a
liquid-liquid or air-liquid interface.  During compression of the membrane,
the surface tension $\gamma$ and area $A$ are measured, from which 
the Gibbs modulus
$E_\mathrm{Gibbs} = \mathrm{d}\gamma/\mathrm{d}\ln A$ is determined. 
The Gibbs modulus $E_\mathrm{Gibbs}$ 
corresponds to the area compression modulus $K_\mathrm{2D}$ in the 
planar trough geometry; we
will show that these two parameters differ substantially in the curved capsule 
geometry.

 In a shear flow rheoscope, a closed capsule is placed
in a liquid phase between two concentric hollow cylinders.  By rotating the
cylinders in opposite directions a shear flow is induced, which deforms the
capsule. Comparing the shape profile with ellipses
gives the compression of the surface and, thus, the 
surface Young modulus \cite{barthes2012microhydrodynamics}.

In a spinning drop apparatus 
a closed capsule is placed in a cylindrical vessel filled
with a fluid. When the vessel is rotated at high frequencies the capsule is
exposed to centrifugal forces, which induce a deformation. Similar to
the  shear flow rheoscope the
surface Young modulus is obtained 
from a shape analysis \cite{pieper1998deformation}.

\begin{figure*}
\centering
\includegraphics[width=0.95\textwidth]{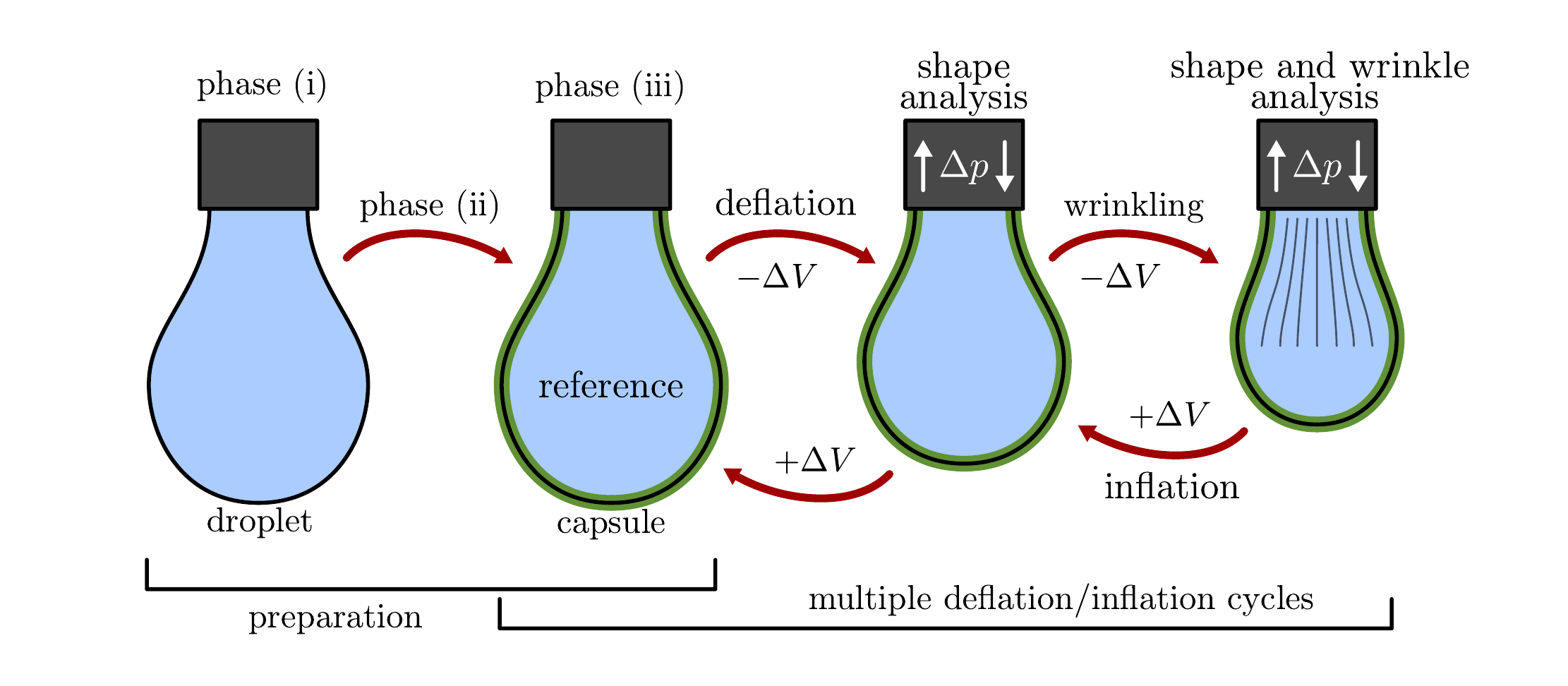}
\caption{
  Typical experimental procedure in  pendant drop elastometry. In a
  first step, a pendant capsule  is prepared by 
  coating a pendant droplet with an elastic shell, 
  for example, by interfacial crosslinking. When the coating process is
  completed, cyclic pressure/volume changes 
are applied to the capsule and images
  are taken continuously during this procedure. 
  Fitting theoretical shapes from the elastic model
  to the shape profile extracted from the image gives the elastic moduli of
  the shell membrane.  
  For sufficiently solid and thin materials wrinkles occur upon
  deflation, which can be analyzed to give the bending modulus of the 
  shell membrane. 
  The procedure allows for multiple inflation-deflation-cycles,
  which can reveal aging effects or hysteresis. 
  Application of different volume change rates can reveal 
   viscoelastic effects.}
\label{fig:1}
\end{figure*}

%%%%%%%%%%%%%%%%%%%%%%%%%%%
\section{Pendant capsule elastometry}

The pendant drop apparatus is widely spread in industrial environments and
research departments.  Typically it is shipped with a software performing a
Laplace-Young analysis on captured images in order to determine the surface
tension of fluid interfaces. 
In this paper, we provide a generalized algorithm as a C/C++ software, 
which is able 
 to perform an analogous shape analysis for elastic membranes in order 
to determine the surface Young modulus $Y_\mathrm{2D}$ 
and the Poisson ratio $\nu_\mathrm{2D}$ (or the Rivlin modulus $Y_M$
and the dimensionless shape parameter $\Psi$)
 of the material.
In section \ref{sec:applications}, we will present 
 examples with several different capsule shell materials,
which demonstrate 
that our software is  widely applicable and
that pendant capsule elastometry 
 results are in good agreement with other rheological measurements. 
 As compared to pendant drop tensiometry, the 
shape analysis of pendant elastic capsules comes at the cost of an
additional amount of runtime (one or few minutes per image), 
but enables the proper characterization 
of  the elastic material properties of capsules.

In the following we will focus on pendant elastic capsules 
 produced by interfacial crosslinking, gelation or polymerization,
 see Fig.\ \ref{fig:1}. 
Consider a droplet of size $\sim 1\mathrm{mm}$ hanging from a
capillary. The inner (liquid) and outer phase (liquid/air) are separated by
a liquid interface with a surface tension 
compensating the pressure difference.  
Surfactants, and potentially crosslinkers, are dissolved
in the droplet or the surrounding fluid.
When forming the droplet, surfactants immediately start
to adsorb to the interface and spread over it.
During equilibration
of bulk and interface surfactant 
concentrations, the surface tension decreases. Though
the interface is now partially occupied by surfactants, 
it is still a liquid
interface obeying the Laplace-Young law. 
This changes when crosslinkers start to
connect previously freely diffusing surfactants 
and turn the
interface into an elastic solid by forming elastic bonds
above a threshold concentration for gelation. 
 After completion of
this crosslinking process, an elastic capsule in its reference, 
i.e., undeformed or stress-free
shape has been formed.

By slowly reducing the volume of the capsule one
observes elastic deformations, which are specific to the microscopic structure
of the membrane.  We neglect such microscopic details by
assuming a homogeneous isotropic material 
and focus on the set of elastic constants,
which describe the macroscopic properties of the membrane.  Nonetheless,
microscopic effects can be observed in the elastic constants,
if these are measured during the course of deflation.
Phase transitions that occur as a function
of the accessible surface area induce a rapid change
in the elastic moduli and are, therefore, detected. 
Viscoelastic or creep behavior are detected, if elastic moduli change 
with the rate of volume reduction.
Aging effects are detected, if elastic moduli change during the course of 
multiple cycles of de- and inflation that are  applied to the capsule.

\begin{figure*}[t]
\centering
\includegraphics[width=0.85\textwidth]{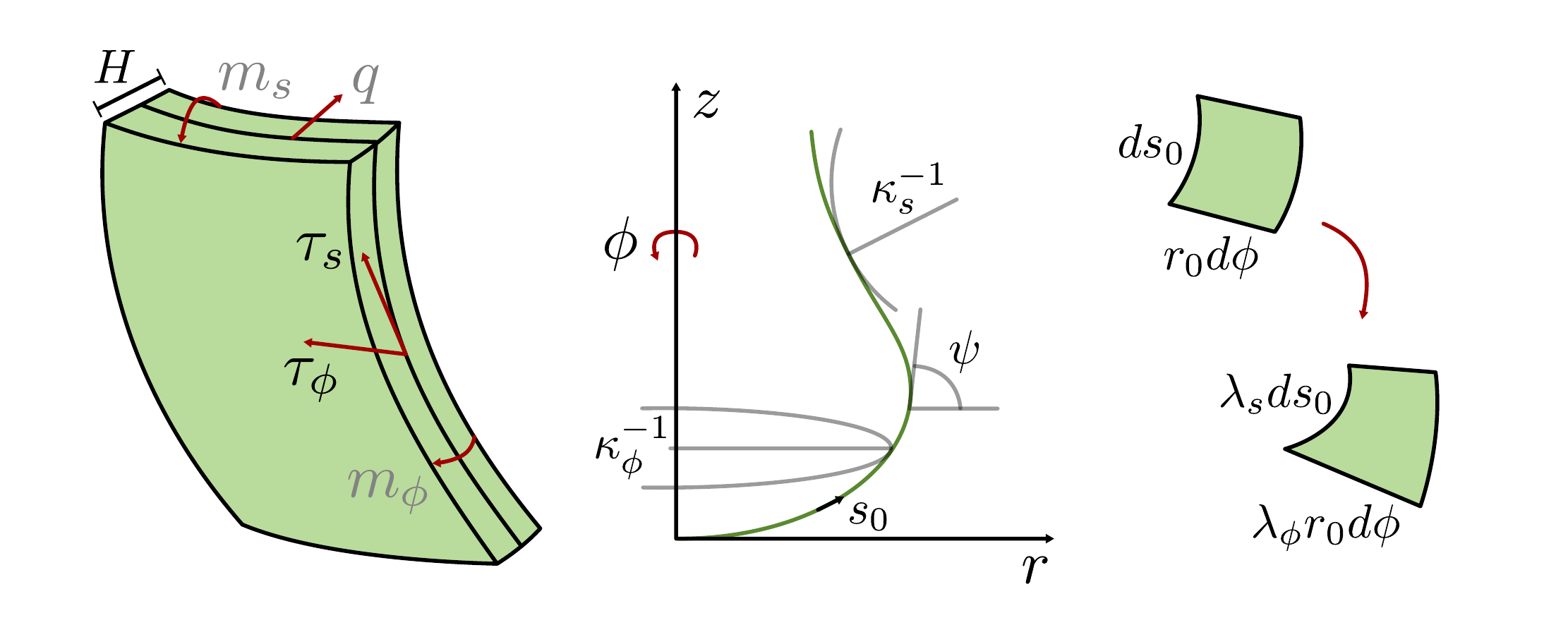}
\caption{Parametrization of the shape equations. 
   Axisymmetric shapes are described by a  shape profile in the
  $r$-$z$-plane, which 
    generates a surface of revolution by rotation with respect to
  the $z$-axis. The shape equations are integrated over the thickness $H$ of
  an infinitesimal membrane patch, which is an approximation for thin
  shells. Forces and torques resulting from curvature are neglected and, thus,
  depicted in gray.
}
\label{fig:2}
\end{figure*}

The general aim of pendant capsule elastometry is the same as in 
pendant drop tensiometry, namely to adjust material parameters 
 (elastic constants or surface tension and, eventually, pressure) 
such that theoretically generated 
axisymmetric capsule/droplet  shapes  optimally fit a given 
experimental shape.
The shape of an elastic capsule can be described by classical elastic shell
 theory (if bending moments are included) 
or  elastic 
membrane theory (if bending moments are neglected) \cite{libai}, 
which requires an elastic material model
specified by its constitutive stress-strain relation or a corresponding 
elastic energy. 
We neglect bending moments and mainly focus on 
 Hookean membrane elasticity throughout this paper
but also discuss  Mooney-Rivlin or neo-Hookean elastic membranes,
which are also implemented in the software.
Each elastic material model is characterized by a set of 
elastic material parameters, such as the 
surface Young modulus $Y_\mathrm{2D}$ 
and the Poisson ratio $\nu_\mathrm{2D}$ in Hookean membrane 
elasticity (or the Rivlin modulus $Y_M$ and the dimensionless
shape parameter $\Psi$ in Mooney-Rivlin membrane elasticity),
which we aim to determine by optimally fitting 
theoretical to experimental shapes.
For an elastic material we also need  an elastically  relaxed
reference shape, with respect to which  local stretch factors 
or strains in the material are defined, in order to 
calculate deformed shapes. This problem will be discussed in 
Sec.\ \ref{sec:reference} in detail.

In a  pendant capsule geometry (see Fig.\ \ref{fig:1})
the capsule is formed hanging from an axisymmetric capillary.
Therefore, we assume axisymmetric shapes
which  can be uniquely described by their shape profile, i.e., 
the intersection of the 
capsule surface with a plane.
We use cylindrical coordinates to describe axisymmetric shapes 
(with the $z$-axis as symmetry axis) and describe the profile 
as contour 
in the $r$-$z$-plane, see Fig.\ \ref{fig:2}. 
The shape profile 
can be trivially acquired experimentally by taking a two-dimensional 
image from the side.

For a given  elastic model, tangential and normal force equilibrium
of stresses and external forces at every point on the 
surface determine the equilibrium shape.
The equations of force equilibrium,  geometric 
relations, and constitutive relations 
 can be used to derive a closed system of  first order
differential equations for $r$, $z$, a slope angle $\psi$ (see
Fig.\ \ref{fig:2}) and the meridional elastic 
stress $\tau_s$ with the arc length $s_0$ of the 
undeformed spherical profile as independent variable, which are called
shape equations and discussed below in Sec.\ \ref{sec:shape_eqs} in 
detail. 
The shape equations use the constitutive relations and are, therefore, 
specific for the elastic model that is used to describe the capsule 
material. 
We discuss shape equations for 
Hookean elasticity, Mooney-Rivlin, and neo-Hookean 
materials in detail in Sec.\ \ref{sec:shape_eqs}.
 The solution of the  shape equations determines the 
theoretical  shape profile of an axisymmetric shape 
for a given set of material parameters,
a given  pressure inside the capsule, and a given elastically relaxed 
reference shape. 
We solve the shape equations  numerically 
by a shooting method (see \ref{sec:shooting}) 
because boundary conditions have to be applied at both ends of the 
shape profile, i.e., at the attachment point to the 
capillary and at the lower apex of the capsule.

%We assume axisymmetric shapes
%since they can be described by a shape profile, 
%which is the intersection of the 
%capsule surface with a plane
%and can be acquired experimentally by taking an image. 

Comparison between experiment and theory
is  achieved by overlaying the theoretical shape and the
image, regardless of the employed elastic model. 
In pendant capsule elastometry 
elastic material parameters of individual capsules are  then 
determined by fitting the parameters (i.e., elastic moduli)
 of the model until the
theoretical shape (for each parameter set obtained by solution 
of the shape equations) optimally 
matches the contour given in the image.
We determine the optimal fit by minimizing the mean-square deviations 
between image contour points and theoretical contour (see
\ref{sec:regression}). 

This approach works with different constitutive relations 
with different numbers of elastic moduli,
as long as the shape is sensitive to each of the employed
material parameters. 
One prominent example for a parameter, 
where the shape is rather insensitive to, is the bending modulus 
$E_B$  for  a thin capsule membrane.
Therefore, the bending modulus  of a thin capsule membrane
cannot be determined reliably by fitting to the shape profiles
but has to be determined via a different approach, namely 
the analysis of wrinkle wavelengths. 
In case of a wrinkling instability, we have to use an effective 
constitutive relation in the wrinkled part of the shape profile
and corresponding effective shape equations, 
which are also discussed in the next Sec.\ \ref{sec:shape_eqs}.

Thus,   for the complete pendant elastometry shape
analysis we have to handle three major tasks, (i) solving the
shape equations for the given elastic model to determine 
the theoretical deformed shape from the elastically relaxed reference shape, 
(ii) decoding the contour from the image and 
(iii) adapting the
model parameters (elastic moduli) to fit the contour.  
Details of the algorithm are described in the Appendix:
(i) solving the shape equations by a multiple shooting method in 
\ref{sec:shooting} and determining the reference 
shape in \ref{sec:app-reference}, (ii)  the 
image processing  in \ref{sec:image}, and (iii) parameter 
determination by  shape regression in \ref{sec:regression}.
In contrast to previous
implementations \cite{knoche2013elastometry}, 
we strongly focus on numerical performance and robustness as
well as a high degree of automation and make the resulting 
C/C++ code publicly available under a GPL License \cite{gplv3}. 
We also implement shape equations for different 
constitutive relations for elastic membranes: nonlinear  Hookean
as in Ref.\ \cite{knoche2013elastometry}, 
strictly linear Hookean as in Ref.\ \cite{Ferri2010,Ferri2012,Cramer2017}, 
Mooney-Rivlin, and neo-Hookean membranes as in Ref.\ \cite{Carvajal2011}.
 We verify our method by analyzing
several materials and comparing the results to rheological measurements.

%%%%%%%%%%%%%%%%%
\section{Shape equations}
\label{sec:shape_eqs}

%%%%%%%%%%%%%5
\subsection{Liquid reference shape}

The shape of a liquid droplet hanging from a capillary can be described by
a system of nonlinear differential shape equations with the 
arc length $s_0$ of the contour as independent variable,
\begin{align}
\begin{split}
r_0^{'}(s_0) &= \cos\psi_0,\hspace*{1em} 
z_0^{'}(s_0) = \sin\psi_0, \\
\psi_0^{'}(s_0) &= (p_0-\Delta\rho g z_0)/\gamma - \sin\psi_0/r_0
\end{split} 
\label{eqn:laplace-young}
\end{align}
(primes denote derivatives $\mathrm{d}/\mathrm{d}s_0$).
The $z_0$-axis is the axis of
symmetry, $r_0$  the radius and $\psi_0$ the
slope angle of the contour.
We use quantities with a subscript ``0'' 
because we will employ a Laplace-Young fit for 
the elastically relaxed 
reference state of our elastic capsule before deformation by 
 volume reduction, see Sec.\ \ref{sec:reference} below.  
The first two equations are geometric relations involving the slope angle
$\psi_0$; the third equation is
 the Laplace-Young force balance equation in cylindrical
parametrization, where we use $\kappa=\mathrm{d}\psi_0/\mathrm{d}s_0$
for the curvature of the droplet. Note that the Laplace-Young shape 
equations \eqref{eqn:laplace-young} are already closed, i.e., 
the right hand side is completely written 
in terms of the three functions $r_0(s_0)$, $z_0(s_0)$, $\psi_0(s_0)$ 
on the left hand side.
 The arc length $s_0$ varies in the range $[0,L_0]$;
 the lower apex is located at $s_0=0$ the drop is attached 
to the capillary at $s_0=L_0$.
The  Laplace-Young shape equations are solved with 
initial conditions $r_0(0)=0$, $\psi_0(0) = 0$, 
and $z_0(0)=\zeta$ arbitrary; the contour length $L_0$ 
is determined by the boundary condition $r_0(L_0) = a/2$, where $a$
is the inner capillary diameter.
The right hand side of $\psi_0'(0)$ at $s_0=0$ is ambiguous
using the initial values;
L'H{\^o}pital's  rule leads to 
$\psi_0'(0)=(p_0-\Delta\rho g \zeta)/2\gamma$ at $s_0=0$ 
which is also needed to start the integration of the shape 
equations.
The solution gives the 
droplet shape $(r_0(s_0),z_0(s_0))$ 
as a function  of the parameters $p_0$, $\Delta\rho$ and
$\gamma$.  The pressure $p_0$ is the hydrostatic pressure at the apex 
(if $\zeta = 0$), $\Delta \rho$
the density difference between the inner and outer phase, 
and $\gamma$ the surface tension.
The pressure difference $\Delta\rho g z_0$ is induced by gravity.

%%%%%%%%%%%%%5
\subsection{Elastic membrane materials}

The Laplace-Young shape equations 
 \eqref{eqn:laplace-young} are well suited for fluid
interfaces, but interfacial crosslinking or gelation
actually turns the interface into a two-dimensional elastic
solid.  From classical shell theory and neglecting bending moments
one derives the elastic shape equations
with the 
arc length $s_0$ of the {\it undeformed} spherical contour 
as independent variable,
\cite{knoche2013elastometry}
\begin{align}
\begin{split}
r^{'}(s_0) &= \lambda_s\cos\psi,\hspace*{1em}
 z^{'}(s_0) = \lambda_s\sin\psi,  \\
\psi^{'}(s_0) &= \frac{\lambda_s}{\tau_s}\left(p-\Delta\rho g z 
 - \frac{\sin\psi}{r}\tau_\phi\right), \\
\tau_s^{'}(s_0) &=  -\lambda_s \frac{\cos\psi}{r} (\tau_s-\tau_\phi)
\end{split}
\label{eq:shape}
\end{align}
(primes denote derivatives $\mathrm{d}/\mathrm{d}s_0$).
The meridional and hoop stretches 
$\lambda_s=\mathrm{d}s/\mathrm{d}s_0$ and $\lambda_\phi=r/r_0$ 
 capture the elastic deformation state 
and are, thus, only  defined with respect to the undeformed 
reference shape $r_0(s_0)$ (with subscript ``0''); 
the corresponding strains are 
$(\lambda_s^2-1)/2\approx \lambda_s-1$ and 
$(\lambda_\phi^2-1)/2\approx\lambda_\phi-1$.
%The principal curvatures $\kappa_s=\mathrm{d}\psi/\mathrm{d}s$ and
%$\kappa_\phi=\sin\psi/r$ derive from differential geometry.
Note that $s$ denotes the arc length of the deformed configuration,
whereas $s_0$ denotes the arc length of the undeformed configuration.
Fig.\ \ref{fig:2} illustrates the involved quantities.
The first to equations are geometric relations involving the slope angle
$\psi$.
The  third and fourth 
 equations describe normal and tangential force balance, respectively.
In the normal force balance, the 
 principal curvatures $\kappa_s=\mathrm{d}\psi/\mathrm{d}s$ and
$\kappa_\phi=\sin\psi/r$ have been used.

It is important to note that eqs.\ (\ref{eq:shape}) are still valid 
regardless  of the constitutive relation. This is also the reason 
why  eqs.\ (\ref{eq:shape}) are not yet closed: we have to 
rearrange 
 the constitutive relations $\tau_s=\tau_s(\lambda_s,\lambda_\phi)$ 
and $\tau_\phi=\tau_\phi(\lambda_s,\lambda_\phi)$ 
in order  to express $\tau_\phi$ and $\lambda_s$ 
on the right hand side of eqs.\ (\ref{eq:shape}) in terms of 
 $\tau_s$ and $\lambda_\phi= r/r_0$, i.e., 
in terms of the functions $\tau_s(s_0)$ and $r(s_0)$ from the left hand side
(and the known reference shape $r_0(s_0)$).
We will discuss closure of the shape equations for different constitutive
relations and also in the presence of wrinkles 
in the following sections.
Once the shape equations (\ref{eq:shape}) are closed, they are 
solved with the boundary conditions  $r(0)=0$, 
$\psi(0) = 0$, and $z(0)=\zeta$ arbitrary at the capsule apex. 
A fourth boundary condition  $\tau_s(0) = \mu$ at the capsule apex
serves as shooting parameter to satisfy  the boundary condition 
$r(L_0) = a/2$ at the capillary (see \ref{sec:shooting}
for the numerical realization of the shooting method).
The right hand sides of $\psi'(0)$ and $\tau_s'(0)$ 
at $s_0=0$ are ambiguous using the initial values;
 L'H{\^o}pital's rule leads to $\lambda_s(0) = \lambda_\phi(0)$  
 and isotropic tensions $\tau_s(0)=\tau_\phi(0)$ at the apex.
 This results in $\tau_s'(0) =0$ and 
 $\psi_0'(0)=\lambda_s(0)(p-\Delta\rho g \zeta)/2\mu$ at $s_0=0$ 
 which are also needed to start the integration.

The pressure $p$ is the hydrostatic pressure at the apex of the deflated shape
(if $\zeta = 0$), 
which is below the pressure $p_0$ of the reference shape, i.e., $p < p_0$.
In principle, information on the pressure $p$ could  be 
 experimentally available if pressure  measurements are possible. 
In the current implementation of the method and all applications
below, the pressure $p$ serves as Lagrange multiplier that is 
changed to control the capsule volume and determined from 
shape fitting along with the  elastic moduli.

%%%%%%%%%%%%%5
\subsection{Nonlinear Hookean elastic membrane}

For a Hookean stretching elasticity 
the meridional and circumferential tensions $\tau_s$ and
$\tau_\phi$ are related to the stretches $\lambda_s$ and $\lambda_\phi$ 
by the constitutive relations
\begin{align}
\begin{split}
\tau_s &=\frac{1}{\lambda_\phi}\frac{Y_\mathrm{2D}}{1-\nu_\mathrm{2D}^2}
    ((\lambda_s-1)+\nu_\mathrm{2D}(\lambda_\phi-1))+\gamma,
\\ \tau_\phi &=
\frac{1}{\lambda_s}\frac{Y_\mathrm{2D}}{1-\nu_\mathrm{2D}^2}
  ((\lambda_\phi-1)+\nu_\mathrm{2D}(\lambda_s -1))+\gamma,
\end{split}
\label{eqn:const-laws}
\end{align}
where 
 $Y_\mathrm{2D}$ is the surface Young modulus 
and $\nu_\mathrm{2D}$ Poisson's ratio. 
Instead of  the surface Young modulus $Y_\mathrm{2D}$ we could 
also use the surface shear modulus $\mu'$ (sometimes called storage modulus 
 $G'$) or the area compression 
modulus $K_\mathrm{2D}$ as alternative elastic constants of the 
membrane material, which are related by 
\begin{equation}
\mu'= \frac{Y_\mathrm{2D}}{2(1+\nu_\mathrm{2D})}
= K_\mathrm{2D}\frac{1-\nu_\mathrm{2D}}{1+\nu_\mathrm{2D}} ~,~
K_\mathrm{2D}= \frac{Y_\mathrm{2D}}{2(1-\nu_\mathrm{2D})}.
\label{eq:muK}
\end{equation}

Although we use a simple Hookean 
elastic energy, the relations (\ref{eqn:const-laws}) are
{\it nonlinear} because 
of the additional $1/\lambda$-factors, 
which arise for purely geometrical 
reasons: the Hookean elastic energy density is defined per 
undeformed unit area, whereas the Cauchy stresses $\tau_s$
and $\tau_\phi$ are defined per deformed unit length.
The relations (\ref{eqn:const-laws})  still contain an interfacial tension 
$\gamma$ because the 
elastic capsule is formed in the initial shape of a fluid interface.
We assume that $\gamma$ is the tension of the fluid interface 
in presence of a saturated interfacial surfactant concentration 
before crosslinking the surfactants to an elastic shell. This
assumption is addressed in detail in Sec.\ \ref{sec:reference}.

The system of shape equations (\ref{eq:shape})
can now be  closed by using on the right hand side 
the constitutive relation for $\tau_\phi$ from eqs.\
(\ref{eqn:const-laws}), 
 and the relation
\begin{align}
\lambda_s &= (1-\nu_\mathrm{2D}^2)\lambda_\phi
  \frac{\tau_s-\gamma}{Y_{2\text{D}}}
  -\nu_\mathrm{2D}(\lambda_\phi-1)+1~~
\mbox{with}~~
\lambda_\phi = \frac{r}{r_0},
\label{eq:Hooke_closure}
\end{align}
which  derives from 
the constitutive relation  for $\tau_s$ from eqs.\
(\ref{eqn:const-laws}).
The resulting shape equations have also been used in 
Ref.\ \cite{knoche2013elastometry}.

%%%%%%%%%%%%%5
\subsection{Strictly linear Hookean, 
Mooney-Rivlin, and neo-Hookean  membranes}

At this point we want to compare to similar approaches
to pendant capsule shapes by shape equations 
in the literature. 
Shape equations very similar  to eqs.\ (\ref{eq:shape}) 
have been obtained in Refs.\ 
\cite{Ferri2010,Carvajal2011,Ferri2012,Cramer2017}, where  the same 
normal and tangential  force balance and geometry relations have been 
employed, however, in combination with   different 
 constitutive relations.
In Ref.\ \cite{Carvajal2011}, an incompressible 
 neo-Hookean constitutive relation 
has been used for the shell material,
 which is a special case of an incompressible  Mooney-Rivlin 
material. In Refs.\ \cite{Ferri2010,Ferri2012,Cramer2017}, 
a strictly linear Hookean 
constitutive law has been used, where the $1/\lambda$-factors
are missing as compared to the relations (\ref{eqn:const-laws}),
(note that constitutive linear Hookean laws in 
Refs.\ \cite{Ferri2012,Cramer2017} contain some misprints).
In Refs.\ \cite{Carvajal2011,Ferri2012}, exemplary 
theoretical shapes 
have been discussed but no elastic parameters have been 
determined from systematically fitting theoretical 
shapes to experimental images, i.e., using a least square
minimization algorithm to optimally match the experimental shape
with a theoretically generated contour.
Therefore, we want to discuss how the shape equations (\ref{eq:shape}) 
can be closed not only for a {\it nonlinear Hookean membrane}
as in (\ref{eq:Hooke_closure}) but also 
for other constitutive relations.

The simplest example is a {\it strictly linear Hookean membrane}, 
where the closure is simply lacking 
one factor $\lambda_\phi$ as compared to eq.\ (\ref{eq:Hooke_closure}) 
\cite{Barthes-Biesel2002}, 
\begin{align}
\lambda_s &= (1-\nu_\mathrm{2D}^2)\frac{\tau_s-\gamma}{Y_{2\text{D}}}
  -\nu_\mathrm{2D}(\lambda_\phi-1)+1~~
\mbox{with}~~
\lambda_\phi = \frac{r}{r_0}.
\label{eq:lin_Hooke_closure}
\end{align}
Thus, the closure relations are, as for the nonlinear Hookean membrane,
analytically accessible.

The {\it Mooney-Rivlin membrane model} is frequently used for polymer 
materials as it  describes membranes made from 
incompressible materials. It describes these materials 
also deep into the nonlinear regime  as it captures effects from 
strain-stiffening.  
It has the constitutive relation 
\cite{Barthes-Biesel2002}
\begin{align}
\begin{split}
\tau_s &=\frac{Y_{M}}{3\lambda_\phi\lambda_s}
    \left( \lambda_s^2- \frac{1}{(\lambda_s\lambda_\phi)^2}\right)
   \left[\Psi+(1-\Psi)\lambda_\phi^2\right]+\gamma,
\\ \tau_\phi &=
\frac{Y_{M}}{3\lambda_\phi\lambda_s}
    \left( \lambda_\phi^2- \frac{1}{(\lambda_s\lambda_\phi)^2}\right)
   \left[\Psi+(1-\Psi)\lambda_s^2\right]+\gamma,
\end{split}
\label{eqn:const-Mooney}
\end{align}
where 
 $Y_M$ is the surface Rivlin modulus 
and $\Psi$ a dimensionless shape parameter. 
A {\it neo-Hookean membrane} has $\Psi=1$. In the limit of small 
stretches a neo-Hookean membrane reduces to a Hookean membrane 
with $Y_{2\text{D}}  =Y_M$ and $\nu_\mathrm{2D}=\nu_\mathrm{3D}=1/2$
(for incompressibility). 
In order to close the shape equations we have to 
use the constitutive relations \eqref{eqn:const-Mooney} to find 
 $\lambda_s$ and $\tau_\phi$  as a function of 
 $\tau_s$ and $\lambda_\phi = {r}/{r_0}$ in order 
to replace $\lambda_s$ and $\tau_\phi$ on the right hand side in the 
shape equations  (\ref{eq:shape}),
as for the Hookean case. Unfortunately, this involves roots of  
fourth order polynomials. Therefore, we perform this task numerically 
in our software.
Note that this numerical solution has to be obtained
in each step of numerical integration of the shape equations, i.e., during
each evaluation of the shape equations, which increases the computational
runtime significantly (roughly by a factor of 10)  as 
compared to fits with the nonlinear Hookean  relation.

%%%%%%%%%%%%%5
\subsection{Wrinkling}
\label{sec:wrinkling}

 The above shape
equations (\ref{eq:shape}) 
only hold for thin materials $H\ll R$, since we neglected bending
elastic energy terms resulting from curvature, which can, in principle, be
included into shape equations (see Ref.\ \cite{Knoche2011a}).
This is justified as the bending modulus is expected to scale 
$E_B\propto H^3$,
whereas Young's modulus scales as $Y_\mathrm{2D}\propto H$. 
 Consequently, for thin capsule shells, 
 the shape profiles are insensitive to changes in the
bending modulus, which makes it practically impossible to infer 
$E_B$ from fitting theoretical shape contours to experimental images.

Nevertheless, we can 
determine the bending modulus in a separate analysis of the
wrinkle wavelength \cite{knoche2013elastometry}.  
Wrinkles in meridional direction are present if $\tau_\phi < 0$, i.e., if
compressive stresses occur in circumferential direction (neglecting 
a small critical Euler stress necessary to trigger wrinkling). 
This condition determines the extent of the wrinkled region in 
meridional direction. 
In order to describe wrinkled shapes violating axisymmetry we use a
pseudo-surface ($\bar{r}(s_0),z(s_0))$ (all modified 
quantities related to the
pseudo-surface are denoted with bars)
representing the average amplitude of the wrinkling modulation.
If $\tau_\phi<0$, the algorithm switches to a different set of shape 
equations for the pseudo-surface which is obtained by explicitly 
setting $\tau_\phi=0$ \cite{knoche2013elastometry}.
The modified set of shape equations is also obtained 
from force-balance for the pseudo-surface.
The meridional stresses for the pseudo-surface are related to the 
original stresses by $\bar{\tau}_s = \tau_s \lambda_\phi/\bar{\lambda}_\phi$,
where $\bar{\lambda}_\phi = \bar{r}/r_0$ is the apparent stretch of the 
pseudo-surface. Together with $\bar{\tau}_\phi = \tau_\phi=0$ we obtain 
shape equations for the pseudo-surface,
\begin{align}
\begin{split}
\bar{r}^{'}(s_0) &= \lambda_s\cos\bar{\psi},\hspace*{1em} 
  z^{'}(s_0) = \lambda_s\sin\psi,  \\
\bar{\psi}^{'}(s_0) &= \frac{\lambda_s}{\bar{\tau_s}}\left(p-\Delta\rho g z 
\right), \\
\bar{\tau}_s^{'}(s_0) &=  -\lambda_s \frac{\cos\bar{\psi}}{\bar{r}} 
  \bar{\tau}_s.
\end{split}
\label{eq:shape_wrinkle}
\end{align}
Note  that these shape equations hold independently of the constitutive
relation of the material. Therefore, they are not yet closed.
To close these shape equations we need to rearrange 
 the constitutive relations $\bar{\tau_s}=\tau_s(\lambda_s,\lambda_\phi)
\lambda_\phi/\bar{\lambda}_\phi$  of the considered model
and the wrinkling condition $0=\tau_\phi(\lambda_s,\lambda_\phi)$ 
in order to 
express $\lambda_s$ 
in terms of $\bar{\tau_s}$ and $\bar{\lambda}_\phi= \bar{r}/r_0$.
We switch to this new set of shape equations (\ref{eq:shape_wrinkle})
 as soon as $\tau_\phi<0$  is reached at $s_0=s_1$ along the contour;
this gives a switching condition that also depends on the 
constitutive relation of the material.  
We switch back to the 
shape equations (\ref{eq:shape}) without wrinkles as soon as 
this condition is violated again at $s_0=s_2>s_1$. 
The extent of the wrinkled region is $L_w = s_2-s_1$.

For a nonlinear Hookean membrane the constitutive relations
(\ref{eqn:const-laws}) lead to a
wrinkling condition 
\begin{align}
\lambda_\phi &=  1
 - \gamma \frac{1-\nu_\mathrm{2D}^2}{Y_\mathrm{2D}}\lambda_s
   -\nu_\mathrm{2D}(\lambda_s -1)
\label{eq:Hooke_closure_wrinkle_condition}
\end{align}
which is also used to identify the wrinkled region $\tau_\phi<0$ 
along the contour.
The constitutive relations 
 (\ref{eqn:const-laws})  also lead 
to the following expression for $\lambda_s$
in terms of $\bar{\tau_s}$ and $\bar{\lambda}_\phi= \bar{r}/r_0$, 
\begin{align}
\lambda_s &= 
\frac{\bar{\lambda}_\phi \bar{\tau}_s+  Y_{2\text{D}} -
  \gamma(1+\nu_\mathrm{2D})}
{Y_{2\text{D}}(1- 2\nu_\mathrm{2D}) - (1-\nu_\mathrm{2D}^2)\gamma^2/Y_{2\text{D}}},
\label{eq:Hooke_closure_wrinkle}
\end{align}
which closes the modified shape equations 
(\ref{eq:shape_wrinkle}) in the wrinkled region.

Similarly we proceed for the constitutive relations of 
a strictly linear Hookean, and eqs.\ (\ref{eqn:const-Mooney}) of 
a Mooney-Rivlin or neo-Hookean membrane in the 
wrinkled region. For the strictly linear Hookean membrane
the wrinkling condition $\tau_\phi=0$ is given by
\begin{align}
\lambda_\phi &=  1
 - \gamma \frac{1-\nu_\mathrm{2D}^2}{Y_\mathrm{2D}}
   -\nu_\mathrm{2D}(\lambda_s -1),
\label{eq:lin_Hooke_closure_wrinkle_condition}
\end{align}
where a factor $\lambda_s$ is missing compared to
\eqref{eq:Hooke_closure_wrinkle_condition}.
Again, we find a relation
  for the meridional stretching factor in terms of $\bar{\tau_s}$ and
  $\bar{\lambda}_\phi= \bar{r}/r_0$,
\begin{align}
&\lambda_s = \frac{Y_\mathrm{2D}(Y_\mathrm{2D} + \gamma(\nu - 1))(1 + 2\nu)}
     {2Y_\mathrm{2D}^2\nu}\nonumber \\
&\pm \frac{\sqrt{Y_\mathrm{2D}^2(Y_\mathrm{2D}^2 + \gamma^2(\nu - 1)^2 +
       2Y_\mathrm{2D}(\gamma(\nu - 1) 
   - 2 \bar{\lambda}_\phi \bar{\tau}_s \nu))}}   {2Y_\mathrm{2D}^2\nu}\,,
\end{align}
where the solution with the negative 
root  has to be chosen.

Unfortunately, for the Mooney-Rivlin membrane analytic expressions
are impracticable since they contain roots of fourth order polynomials. 
However, $\lambda_s$ and $\lambda_\phi$ can be reliably determined 
by numerically solving
\begin{align}
\tau_\phi(\lambda_s, \lambda_\phi) = 0 \hspace*{1em} \text{and} \hspace*{1em}
\bar{\tau}_s - \tau_s(\lambda_s, \lambda_\phi) 
  \lambda_\phi / \bar{\lambda}_\phi = 0.
\end{align}
Note that, in the wrinkled region, $\lambda_s$ and $\lambda_\phi$ have to be
determined, whereas in the non-wrinkled region $\lambda_s$ and $\tau_\phi$
have to be determined. The solution $(\lambda_s, \lambda_\phi)$ of the above
set of equations closes the shape equations \eqref{eq:shape_wrinkle} and can,
in principle, be obtained in the same way for any constitutive law.

The extent of the wrinkled region where $\tau_\phi < 0$ of course depends 
on the value of the interfacial tension $\gamma$ in  all 
constitutive relations (\ref{eqn:const-laws}) or 
(\ref{eqn:const-Mooney}). The fact that we generally
obtain good agreement with experiments regarding the extent 
of the wrinkled region also supports the inclusion of the 
interfacial tension into the constitutive relations.

%%%%%%%%%%%%%%%%%%%%%%%%%%%
\section{Equilibrium and reference shapes}
\label{sec:reference}

\begin{figure*}[t]
\centering
\includegraphics[width=\linewidth]{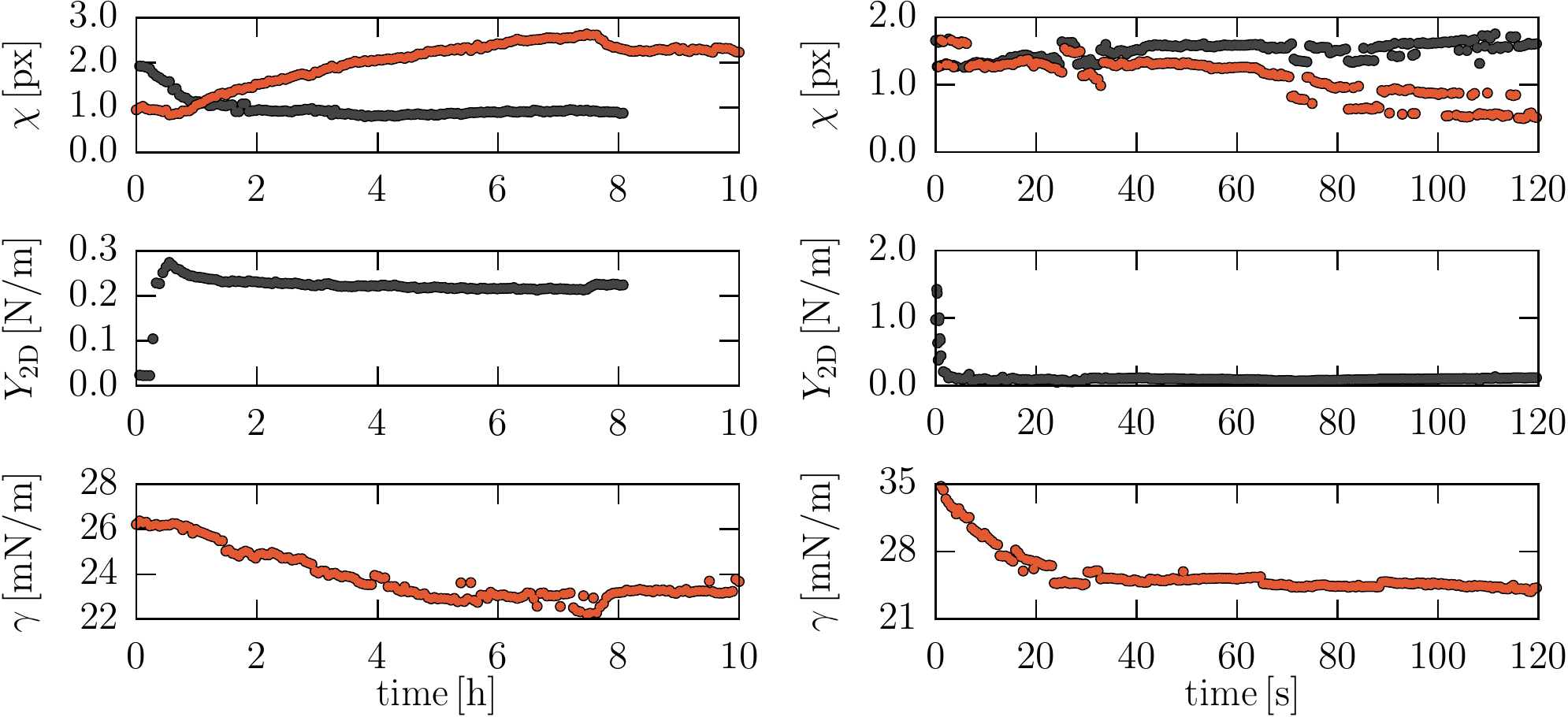}
\caption{
  The interfacial gelation phase (ii) fitted with a model for liquid interfaces
  (Laplace-Young, orange) and a model for 
  elastic interfaces (Hooke membrane, gray), where we 
  use the reference shape from the very beginning of phase (ii), 
  i.e., phase (i). No active pressure or volume change have been applied.
  \textbf{Left:} $\mathrm{H}_2\mathrm{0}$-droplet coated with
  $\mathrm{Ce}(\mathrm{SO}_4)_2$ and $\mathrm{CTAB}$-surfactants in a dodecane phase.  
  \textbf{Right:} 
  OTS-capsule, i.e., a \emph{p}-xylol droplet in solution 
  with 1,2,4-trichlorobenzene and coated
  with OTS in a glycerol-water mixture.
}
\label{fig:network-formation}
\end{figure*}

Solutions of the elastic shape equations (\ref{eq:shape}) 
presume an elastically relaxed reference shape,
with respect to which elastic strains are defined.
The choice of the reference shape is subject to certain assumptions which 
will be discussed in this section.

Capsule formation by crosslinking or polymerization proceeds
via three phases (see Fig.\ \ref{fig:1}).
In phase (i) we have a liquid drop without any surfactants
and a stationary shape (which is a teardrop shape due to gravity).
Phase (ii) starts when surfactants and/or crosslinkers are 
added to one of the bulk phases such that adsorption of surfactants
and subsequent crosslinking into a two-dimensional network can occur.
If surfactants and/or crosslinkers are dissolved in the droplet
or the surrounding fluid, then phase (ii) starts immediately when
forming the droplet.
In phase (ii) the shape changes and the capsule 
finally  reaches a new equilibrium shape. 
In phase (iii) the capsule is in its new stationary shape after 
successful crosslinking; this is the state where the deflation 
experiment is started.

For the regression of deflated capsule shapes we usually assume 
that the equilibrium state reached in phase (iii) is 
identical to the elastically relaxed state and, thus, free 
of elastic tensions. 
As discussed above, we also assume in the constitutive relations 
 \eqref{eqn:const-laws} that 
the surface tension $\gamma$ gives a
constant contribution to the tensions $\tau_s$ and $\tau_\phi$.
Then the surface stress in the elastically relaxed state of the membrane is
solely determined by the isotropic surface tension $\gamma$.
We thus assume in \eqref{eqn:const-laws} that 
the elastically relaxed state 
can be described as a liquid drop shape using the  Laplace-Young 
equation. Based on these two assumptions
 we  use a Laplace-Young fit for  the equilibrium shape in phase (iii).

These two assumptions are based on the 
following picture for the crosslinking process in phase (ii):
  When adding
surfactants to one of the bulk phases at the beginning of phase (ii),
 the surface tension typically decreases
linearly or exponentially in time until it reaches a plateau 
 at the equilibrium surface tension $\gamma_A$. 
The actual crosslinking of the membrane only happens 
  {\it after} the  plateau at the surface 
tension $\gamma_A$ has been reached. During crosslinking 
the interfacial tension  $\gamma_A$   of the fluid interface 
remains unchanged. 
If this picture is valid, we should observe a sagging of the capsule 
under the action of gravity while 
a decreasing surface tension gives shapes that can be successfully fitted 
using the Laplace-Young shape equations (\ref{eqn:laplace-young}).
The sagging should stop when the surface tension reaches the plateau.
During this plateau phase the crosslinking is established, while 
the capsule shape is unchanged. 
Fig.\  \ref{fig:network-formation} (right) shows an example of an OTS-capsule 
where all these features can indeed be observed. Fitting the 
shape using the  Laplace-Young shape equations gives only 
small errors and the interfacial tension $\gamma$ follows the expected 
temporal evolution.

There are, however, capsule formation processes which deviate 
from this picture. 
Another possible scenario is that the formation of a solid shell 
by crosslinking happens earlier in phase (ii) but further polymerization 
during phase (ii) 
generates  elastic strains and stresses.
All further shape changes during phase (ii) have to be interpreted as a
result of strain and stress generation during the polymerization process,
and the capsule shell is 
pre-stressed in  the equilibrium state 
in phase (iii). Then the elastic reference shape is not 
exactly known and, in principle, 
 can  be any of the shapes encountered in phase (ii). 
One extreme  assumption is 
 that crosslinking is fast and 
 a  solid membrane is established right at the 
beginning of phase (ii). Then the shape in the beginning of phase (ii) directly
after addition of surfactants and crosslinkers 
can be viewed as the elastic
 reference shape and all subsequent shapes
should be fitted with an elastic model using this reference shape. 
Fits with the elastic model should reveal how strains, stress, and 
elastic moduli evolve during phase (ii).

In order to decide which choice of reference shape 
 is most appropriate, one can 
try different fits using different shapes from phase (ii) 
as elastically relaxed reference shapes
(for example, from the end or the beginning of phase (ii)).
All shapes before the reference shape are fitted using the 
Laplace-Young shape equations and described by an 
interfacial tension $\gamma$ that decreases in time.
All shapes following the reference 
shape are fitted using the elastic shape equations and described 
by a surface Young modulus $Y_\mathrm{2D}$ and a Poisson ratio
$\nu_\mathrm{2D}$, which evolve in time. 
The reference shape giving the best fits (with 
smallest errors) should be chosen. 
Moreover, choices of reference states 
 producing unphysical results, such as 
a surface Young modulus $Y_\mathrm{2D}$ which is decreasing in time 
during  the crosslinking process in phase (ii) 
(more crosslinks or junction points should 
always increase  $Y_\mathrm{2D}$),
should be discarded. 

\begin{figure*}[t!]
\centering
\includegraphics[width=0.9\linewidth]{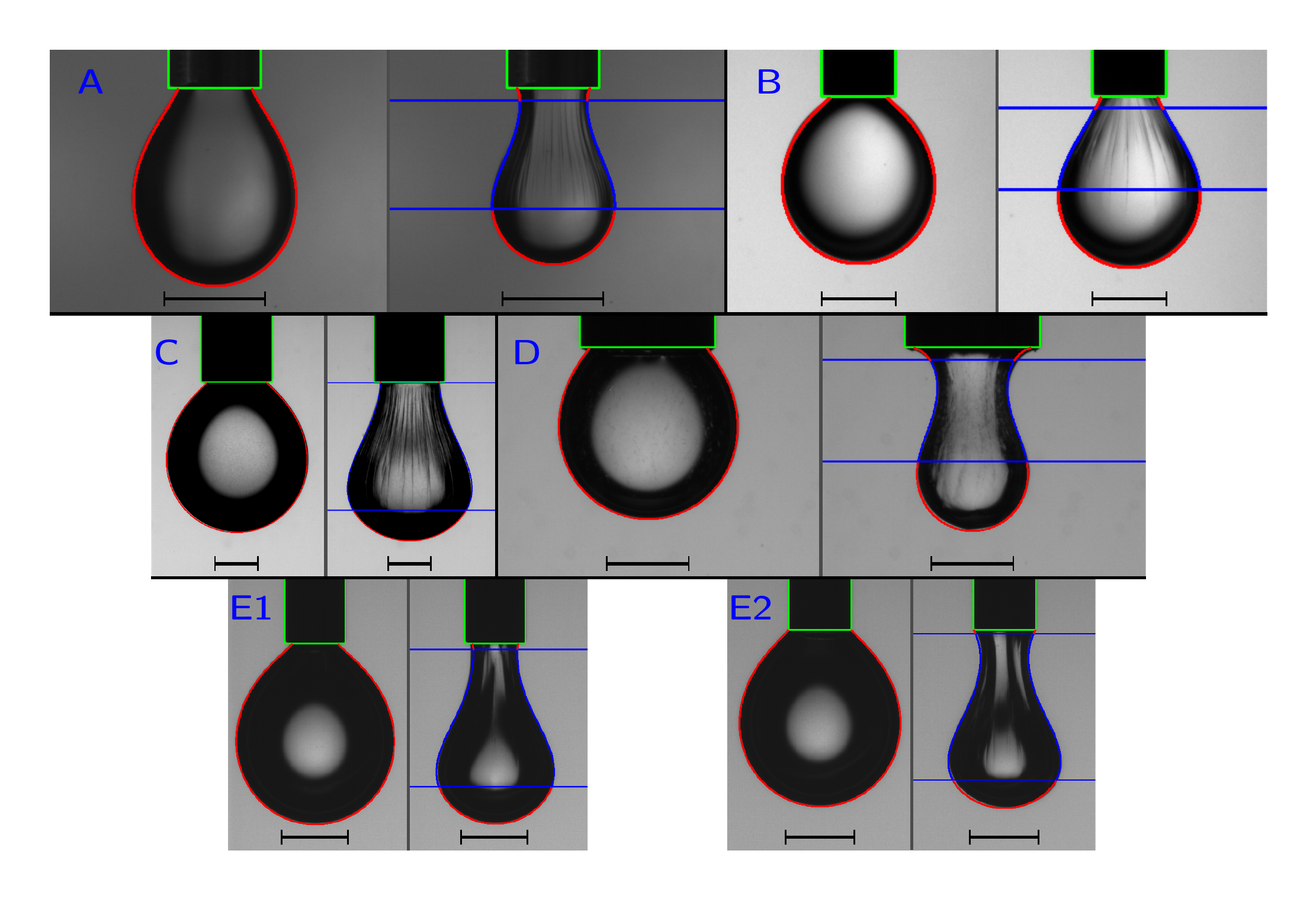}
\caption{
  Pendant capsule elastometry results for five  different 
  types of capsules. The above images are the original input files
  overlayed with the best fit theoretical contour, the scale-bar
  is of length $1\,\mathrm{mm}$. These images are automatically generated
  by our Software.
  \textbf{\textsf{A)}} Reference and deflated shape of a
   FC-40 droplet coated with supramolecular polymers and crosslinked with 
   cucurbit[8]uril in
  $\mathrm{H}_2\mathrm{O}$ \cite{Salmon2016}.  Both the capsule contour and
  the wrinkling region are perfectly described by the elastic model. 
  The Laplace-Young
  analysis yields a surface tension $\gamma = 20.0\,\mathrm{mN/m}$.  The elastic
  analysis yields an area compression modulus $K_\mathrm{2D}=44\,\mathrm{mN/m}$,
  a Poisson ratio $\nu_\mathrm{2D}=0.29$, a bending modulus $E_B = 5\cdot
  10^{-16}\,\mathrm{Nm}$ and a layer thickness $H=256\,\mathrm{nm}$.  
  \textbf{\textsf{B)}} Reference and deflated shape of 
  a dodecane droplet coated with three layers of PMAA/PVP in
  $\mathrm{H}_2\mathrm{O}$ \cite{le2014interplay}.  The Laplace-Young analysis
  yields a surface tension $\gamma = 13.2\,\mathrm{mN/m}$.  The elastic analysis
  yields an area compression modulus $K_\mathrm{2D}=141\,\mathrm{mN/m}$, a
  Poisson ratio $\nu_\mathrm{2D}=0.75$, a bending modulus $E_B=2.20\cdot
  10^{-14}\,\mathrm{Nm}$ and a layer thickness $H=1.28\,\mu \mathrm{m}$.
  \textbf{\textsf{C)}} Reference and deflated shape of an OTS-capsule, 
  i.e., \emph{p}-xylol droplet in solution with 1,2,4-trichlorobenzene 
 and coated
  with OTS in a glycerol-water mixture, 
  see also Fig.\ \ref{fig:ots},
  \textbf{\textsf{D)}} Reference and deflated shape of a Span 65 capsule, 
  i.e., $\mathrm{H}_2\mathrm{0}$-droplet coated with Span 65 (sorbitan
  tristearate) in dodecane, see also Fig.\ \ref{fig:span} (left).
  \textbf{\textsf{E)}} Reference and deflated shape of an amino-functionalized 
  polyacrylamide capsule, i.e., $\mathrm{H}_2\mathrm{0}$-droplet with
  $\mathrm{Na}_2\mathrm{CO}_3$,
  $\rm{N}\text{-}(3\text{-}\rm{Aminopropyl})\text{-}\rm{methacrylamide}$,
  and DTAB \textbf{\textsf{E1)}} or CTAB \textbf{\textsf{E2)}} surfactants, 
  surrounded by an outer phase with \emph{p}-xylol
  and sebacoyl dichloride, see also Fig.\ \ref{fig:span} (right). 
  For all five capsule types predicted wrinkle regions (blue lines)
  fit the actual wrinkled area quite well. 
  The wrinkles of the Span 65
  capsule are hardly visible by eye, probably because of a very thin shell 
  and, thus, a small wrinkle wavelength. Span 65 is expected to form molecular
  monolayers, which is consistent with this interpretation.
}
\label{fig:figs}
\end{figure*}

Two examples are shown in Fig.\  \ref{fig:network-formation}.
OTS-capsules show the typical sagging in phase (ii) and 
can be fitted quite well 
with the Laplace-Young shape equations giving a 
surface tension $\gamma$, which at first
decreases linearly or exponentially and then reaches a plateau, consistent with 
the standard scenario that the shell is crosslinked at the end of phase (ii). 
But the shapes can also be fitted quite well assuming that crosslinking 
is established at the beginning of phase (ii); then the observed  
sagging leads to fits with a decreasing Young modulus $Y_\mathrm{2D}$
and should, therefore, be discarded as unphysical.

The second, untypical 
example are $\mathrm{H}_2\mathrm{0}$-droplets coated with
coagulated films of $\mathrm{Ce}(\mathrm{SO}_4)_2$ 
and $\mathrm{CTAB}$-surfactants, 
which only show little sagging during crosslinking and 
even develop wrinkles already during phase (ii), which is a strong 
hint that a solid membrane had been established early in phase (ii). 
The crosslinking process is much slower for these capsules. 
Here, fits with the Laplace-Young shape equations give a 
decreasing  $\gamma$ that reaches a plateau; the resulting 
fit errors are, however, quite large and growing in time.
The assumption that the shape in the beginning of phase (ii) is 
already crosslinked and can be regarded as the elastically relaxed reference
shape gives a Young modulus $Y_\mathrm{2D}$, which increases 
sharply in the beginning of phase (ii) and then reaches a plateau;
there is no pronounced decrease in  $Y_\mathrm{2D}$. Fit errors 
for this scenario are decreasing in time. 
 The fit errors for the two fitting approaches actually 
show an intersection point early in phase (ii). 
Between the beginning of phase (ii) and the intersection point,
the capsule shape is adequately described by the liquid model. Beyond the
intersection point the elastic model provides a more accurate description than
the liquid model. 
 One might conclude that the formation of the network at
the interface is completed, when the system passes the intersection point.
The surface Young modulus does not change significantly after passing the
intersection point, which confirms our conclusion. 
Comparing fit errors could serve as  a simple
method to estimate 
 the time needed to built a crosslinked solid shell
 for different materials
or chemical processes.

%%%%%%%%%%%%%%%%%%%%%%%
\section{Software overview}

The software and source code \cite{opencapsule2017}
provided with this paper are freely available at 
\texttt{github.com} under a GPL license \cite{gplv3}.
It is a command line program developed in C/C++ and most compatible with
Linux/Unix.  Usage is fairly simple and a guideline (README.md) is provided
as part of the github repository.
We give a brief description of how the program
works and how the typical workflow looks like. 

Presuming that at least one image of the reference capsule and at least one
image of a deformed capsule is given, a first call \texttt{OpenCapsule} will
establish the workspace, i.e., create folders for the input/output files as
well as a standard configuration file. The essential information in the
configuration file should be updated according to the needs. In particular,
the density difference $\Delta \rho$ between the inner and the outer phase is
needed, as well as the outer capillary diameter $b$.  
 Both can be
  manipulated via the corresponding environment variables
  \texttt{EXPERIMENT\_DENSITY} and \texttt{EXPERIMENT\_CAPDIAMETER}.  In
  addition, the names and  paths of the image files need to be
  specified. Files have to be listed (separated by colons) next to the
  environment variables \texttt{REFERENCE\_SHAPE} and \texttt{ELASTIC\_SHAPE}.
  Note that the software searches for images by default in the
  \texttt{./input/}-folder.  If images are placed somewhere else, the path
  should be specified via the variable \texttt{INPUT\_FOLDER}. Requirements
  for capsule images are detailed in the appendix.

This suffices to run the first analysis. To check if everything
works correctly the command \texttt{OpenCapsule -r} should be
called, which will analyze the reference shapes and determine the surface
tension as an average over all given images and, of course, for each
individual image. This analysis can also be used to fit the 
deformed shapes with the Laplace-Young equation, e.g. to determine
the Gibbs-modulus. If the results are satisfactory, 
the command \texttt{OpenCapsule -s} will run the elastic analysis. 
Both types of analyses are completely automatized. 
The
essential numerical results are placed in the \texttt{./global\_out/}-folder.
The results for the reference shapes are listed in \texttt{reference.dat};
the results for the deformed shapes in \texttt{sequence.dat}.
Though no graphical user interface is provided, the results will be printed in
a comprehensive html-report, which can be opened in a web browser.  
This report contains the original capsule images
with an overlay of the theoretical shape and a scale bar
(see Fig.\ \ref{fig:figs}),
from which one can instantly judge if the fitting procedure was successful.
In case of failure, one should adapt the configuration 
file according to the guideline.
Setting up a proper configuration file once for a specific capsule type is
typically sufficient. Afterwards it can be used without changes for the same
type of capsules.

%%%%%%%%%%%%%%%%%%%%%%%%%
\section{Reference analysis}
\label{sec:reference_app}

\begin{figure*}[t!]
\centering
\includegraphics[width=\linewidth]{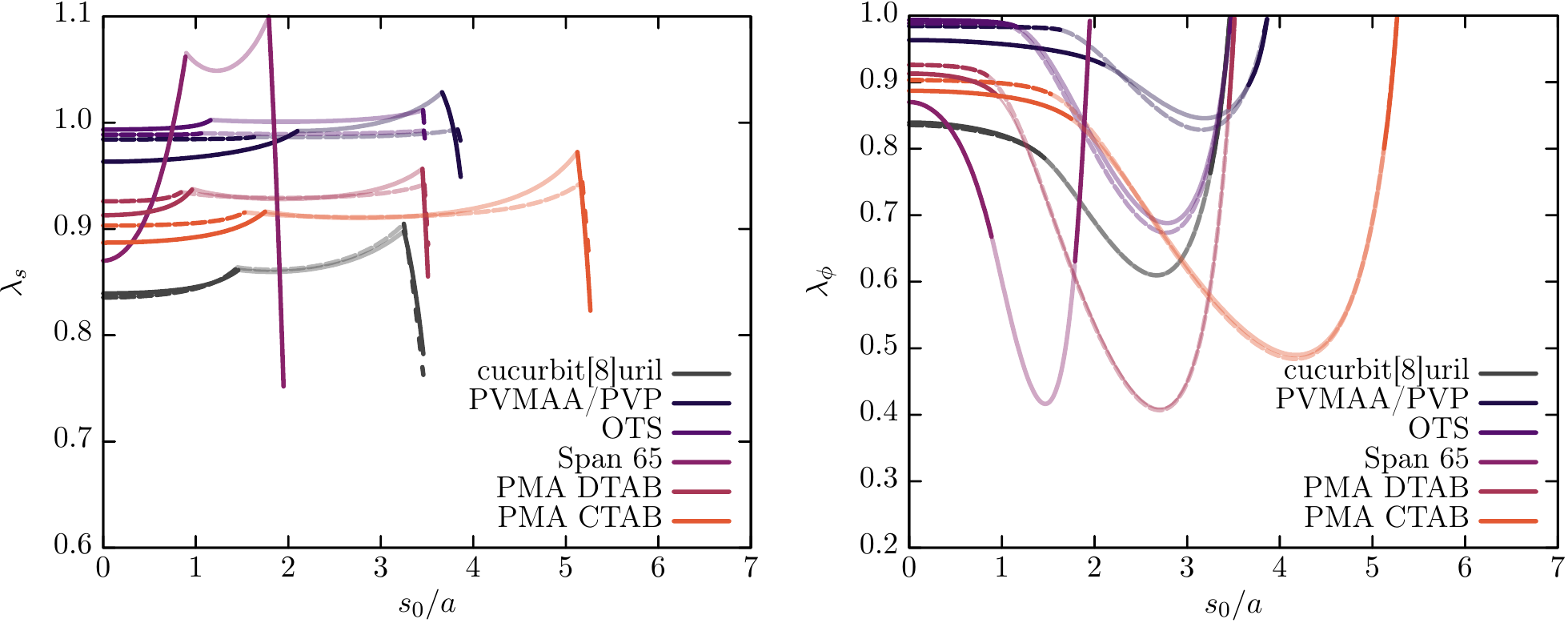}
\caption{ 
  Stretching factors $\lambda_{s,\phi}$ along 
    the deflated capsule contours
    from Fig.\ \ref{fig:figs} obtained from shape regressions with the
    nonlinear Hookean elasticity (solid lines), and the Mooney-Rivlin
    or neo-Hookean elasticity model (dashed lines).  The wrinkled
    region is indicated by transparent lines (either solid or dashed) for both
    elasticity models.
    \textbf{Left:} Meridional stretching factor
    $\lambda_s$ as a function of the  arc length $s_0$.  Except of the
    cucurbit[8]uril and the Span 65 capsules, all capsules have small strains 
    $\vert
    \lambda_s - 1\vert < 0.1$ and, thus, can be successfully treated with
    small strain approximations, i.e., the nonlinear Hookean elasticity model.
    \textbf{Right:} Circumferential stretching factor $\lambda_\phi$ 
    (apparent stretch $\bar{\lambda}_\phi$ in the wrinkled region) as a 
    function of  $s_0$. 
   Particularly in the wrinkled region,
    we see significant deviations from the small strain limit $\vert
    \lambda_\phi - 1\vert<0.1$.  In the wrinkled region we 
    use, however, 
   a different constitutive law ($\tau_\phi = 0$) independent of
   $\bar{\lambda}_\phi$. In
    the non-wrinkled region, we find again small strains $\vert
    \lambda_\phi - 1\vert < 0.1$ for PVMAA/PVP, OTS, and polyacrylamide
    capsules. 
}
\label{fig:lambdas}
\end{figure*}

The shape analysis is split into two batched parts, 
which are (i) reference shape analysis 
and (ii) deformed shape analysis.
For the former one it is
advantageous to analyze as many images as possible showing the same,
undeformed state of the capsule. This is particularly necessary if 
images are slightly blurred from camera shake or capsule motion.
Averaging over all images improves accuracy,
which is important here, since we use the reference shape and parameters
during the complete analysis of the deformed shapes. The reference analysis
gives the surface tension $\gamma$ and the shape profile that 
is necessary to define the
strains in the elastic analysis.

From the experiment we know the outer
capillary diameter $b$ and the density difference $\Delta \rho$ both in SI
units. Solutions of the shape equations have to match the inner capillary of
width $a$, which is the relevant length scale. 
This quantity is typically specified by the needle manufacturer,
but due to material sediments, which potentially change the effective
inner capillary diameter, we prefer to measure it directly from the image.
Actually, we determine $\alpha$ in units of the image (pixels) 
by choosing it as an independent fit parameter that
rescales all lengths occuring in the Laplace-Young shape equations.
From image processing we also know the outer capillary
diameter $\beta$ in units of the image (pixels).
Thus, we find $a = \langle \alpha /\beta \rangle b$, which
is the effective inner capillary diameter in SI units. 
Scaling dimensionless lengths
with $a$ transforms them to SI units.  We introduce dimensionless quantities
$\tilde{p}_0 = a p_0 /\gamma$ and 
$\Delta \tilde{\rho} = a^2 \Delta \rho g
/\gamma$ and minimize the mean square deviation between shape
and contour with respect to the parameter set $\vec{x}_0 =
(\tilde{p}_0, \Delta \tilde{\rho}, \alpha)$.  After successful
minimization, we obtain the surface tension via $\gamma = a\Delta\rho g /
\Delta \tilde{\rho}$. To prepare all contours for the
elastic analysis we scale them with $1/\alpha$ 
and thereby transfer them to dimensionless units.

%%%%%%%%%%%%%%%%%%%%%%%%
\section{Elastic analysis}
\label{sec:elastic_analysis}

In the elastic regression we determine the area compression modulus
$K_\mathrm{2D}$ and the Poisson ratio $\nu_\mathrm{2D}$ by minimizing the
mean-square deviations between image contour points and theoretical contour,
i.e., with respect to the parameter set $\vec{x}=(\tilde{p}, \nu_\mathrm{2D},
\tilde{K}_\mathrm{2D})$, where $\tilde{p} = a p /\gamma$ and
$\tilde{K}_\mathrm{2D} = K_\mathrm{2D}/\gamma$.  From these quantities we also
obtain the surface Young modulus $Y_\mathrm{2D} =
2K_\mathrm{2D}(1-\nu_\mathrm{2D})$, see eq.\ (\ref{eq:muK}).  
For the
  Mooney-Rivlin elasticity model we determine analogously the parameter set
  $\vec{x} = (\tilde{p}, \Psi, Y_M)$ within the shape regression. For the
  neo-Hookean elasticity model we keep $\Psi = 1$ fixed during the shape
  regression.

It is not required that elastic shapes are ordered chronologically,
but it decreases the runtime significantly, since the final parameters of
a deformed shape can be used as an initial guess for the following shape,
which is probably deformed by a similar extent. 

In the current implementation, the pressure $p$ is an 
additional fit parameter and will also be determined from 
fitting calculated shapes to an image. 
In this implementation the elastometry method also serves 
as pressure measurement and  no additional pressure 
measurement is necessary.
If such information is experimentally available from additional
measurements, it could be 
used to improve the results for the elastic moduli,
by fixing the pressure to the experimentally obtained value.

After a successful regression we can estimate the bending
modulus \cite{knoche2013elastometry}
\begin{equation}
  E_B = \Lambda^4 \bar{\tau}_s/16\pi^2L_w^2
\label{eq:wrinkle}
\end{equation}
 by an image analysis of
the wrinkles. We determine the length $L_w$ of the wrinkles
in meridional direction directly from the
shape by finding the zero crossings $s_1$ and $s_2$ 
of $\tau_\phi(s_0)$. The interval
$[s_1,s_2]$ in which wrinkles occur sets the meridional extent 
of the wrinkled region, in which we 
determine the average meridional tension
$\bar{\tau}_s$. 
It is important to note that the extent of the wrinkled region is 
not a fit parameter but is predicted by the shape equations and 
the wrinkling criterion $\tau_\phi<0$, according to Sec.\ \ref{sec:wrinkling}.
In order 
to estimate the wrinkling wavelength $\Lambda$, we select the
wrinkling region from the image and perform an edge detection with high
sensitivity and only in horizontal direction to estimate the average distance
between the wrinkles, which should correspond to one wavelength $\Lambda$.
We then count the detected edge pixels $N_E$ and the total number $N$ of
pixels in the selected region. The ratio $N/N_E$ now approximately gives the
number of wrinkles apparent in the image, if we consider the region as
rectangular with equidistant vertical wrinkles.  Finally, we use the
estimate $\Lambda \approx \pi\bar{r}N_E/N$, 
where $\bar{r}$ is the capsule radius $r(s_0)$ averaged over the interval
$[s_1,s_2]$.

Knowing $Y_\mathrm{2D}$,
$\nu_\mathrm{2D}$ and $E_B$, 
we can directly determine the 
F\"{o}ppl von K\'{a}rm\'{a}n number 
$\gamma_\mathrm{FvK} = Y_\mathrm{2D}R_0^2/E_B$ of the capsule, 
where $R_0$ is the maximum capsule radius.
If we assume that the shell material is a thin layer of a 
three-dimensional isotropic elastic material, 
we additionally find the thickness 
$H =[12E_B(1-\nu_\mathrm{2D}^2)/Y_\mathrm{2D})]^{1/2}$ of the layer,
where we use $Y_{\rm 2D} = Y_{\rm 3D}H$, 
 $E_B =  Y_{\rm 3D}H^3/12(1-\nu_\mathrm{3D}^2)$ and $\nu_{3D} = \nu_\mathrm{2D}$.

%%%%%%%%%%%%%%%%%%%%%%%
\section{Applications}
\label{sec:applications}

In this section we demonstrate the wide applicability of 
our pendant capsule elastometry software to different 
capsule materials,  see Fig.\ \ref{fig:figs}. 
We apply our software to crosslinked polymeric capsule shell 
   materials like OTS
(octadecyltrichorosilane) (Fig.\ \ref{fig:figs} C)
 and  amino functionalized polyacrylamide (Fig.\ \ref{fig:figs} E), 
 as well as more exotic capsule materials like Span 65
\cite{torcello2011surface,krishnaswamy2007nonlinear,erni2005sorbitan}, which
is a food emulsifier (Fig.\ \ref{fig:figs} D). 
Moreover, Fig.\ \ref{fig:figs} shows 
  analyses of two sorts of capsules from 
  literature, cucurbit[8]uril-capsules 
that have been introduced and discussed  in Ref.\ \cite{Salmon2016}
(Fig.\ \ref{fig:figs} A) 
and  PMAA/PVP-capsules  from Ref.\ \cite{le2014interplay} 
(Fig.\ \ref{fig:figs} B).
In addition, the method has been used previously 
(in a less advanced implementation)
on hydrophobin-coated 
air bubbles \cite{knoche2013elastometry}.

As can be seen in Fig.\ \ref{fig:figs}, nonlinear Hookean fits for all capsule
materials work well and correctly predict the extents of the wrinkled regions
(blue lines).  The different capsule materials that could be analyzed have
quite diverse area compression moduli ranging from 
$K_\mathrm{2D} \sim 50 \,\mathrm{mN/m}$ 
(polyacrylamide capsules and cucurbit[8]uril-crosslinked capsules) 
to $K_\mathrm{2D} \sim 4\,\mathrm{N/m}$ (OTS-capsules), which
corresponds to two orders of magnitude.  The bending moduli from the wrinkle
analysis range from $E_B = 5\cdot 10^{-16}\,\mathrm{Nm}$
(cucurbit[8]uril-crosslinked capsules) to $E_B = 10^{-13}\, \mathrm{Nm}$
(OTS-capsules from Fig.\ \ref{fig:multi}).  
For the Span 65 capsules we find
even lower bending moduli of order $E_B = 10^{-20}\,\mathrm{Nm}$ assuming a
quite short wrinkle wavelengths just below the image resolution.  However, the
existence of these wrinkles could not be verified experimentally, yet.

Capsules in Fig.\ \ref{fig:figs}  develop a ``neck'' upon deflation.
We note that this neck is not associated with any mechanical 
instability (e.g., a buckling-type instability
\cite{Knoche2011a,Knoche2014a}),
i.e., there is no bifurcation between different types of shapes 
upon deflation but all shapes continuously evolve into the 
necked shapes. 
The deflated shapes exhibit high compressional stretches in particular 
in the wrinkled region as Fig.\ \ref{fig:lambdas} shows, where the
resulting  stretches $\lambda_s$ and $\lambda_\phi$ are plotted along the 
deflated contours for all capsules  shown in  Fig.\ \ref{fig:figs}
and, indeed,  
values significantly smaller than 1 occur for $\lambda_\phi$. 
This raises the question whether nonlinear effects are adequately treated
by the nonlinear Hookean material model. This model contains nonlinearities 
only via the  $1/\lambda$-factors in the constitutive relations
(\ref{eqn:const-laws}), which arise because we use Cauchy stresses
defined per deformed unit length in the force-equilibria.

Our model explicitly includes, however,  wrinkle formation, 
which is a nonlinear phenomenon. 
We actually use a different constitutive relation $\tau_\phi=0$ 
(see eqs.\ (\ref{eq:shape_wrinkle}) above) in the wrinkled region,
such that a further decrease in the  apparent stretch 
$\bar{\lambda}_\phi= \bar{r}/r_0$ in the wrinkled region does no longer 
lead to increased compressive stresses $\tau_\phi$ but only 
modifies the effective constitutive law  (\ref{eq:Hooke_closure_wrinkle})
for the meridional stresses $\bar{\tau}_s$ of the pseudo-surface. 
Therefore, small values of $\bar{\lambda}_\phi$ in the wrinkled region 
do {\it not} imply that a different, more appropriate 
nonlinear  constitutive relation should be used. The results 
in Fig.\ \ref{fig:lambdas}  show that strains  $|\lambda_\phi-1|$
 become large only in this wrinkled region. 
This suggests that usage of the nonlinear Hookean elasticity 
 is justified.

In order to investigate nonlinear effects further, we also performed
fits of all capsules from Fig.\ \ref{fig:figs} with a 
  nonlinear  Mooney-Rivlin or neo-Hookean elasticity model. 
  The resulting best fits for the  theoretical contour
  are not distinguishable from the nonlinear Hookean contours 
   shown in Fig.\ \ref{fig:figs}, which already suggests that 
  nonlinearities are already adequately treated by the nonlinear
  Hooke law. 
In Fig.\ \ref{fig:lambdas} we also 
   compare the stretches resulting from shape
  regressions  with the
  nonlinear Hookean and the Mooney-Rivlin or neo-Hookean elasticity model
 for all deflated capsule shapes from Fig.\ \ref{fig:figs}. We
  see that strains are similar for both models
  for any of the capsule materials, except of the Span 65 capsules, where we
  were not able to obtain a reliable fit result using the Mooney-Rivlin
  elasticity model.  For Span 65 the meridional stretching factor varies
  strongly with the arc length indicating strongly inhomogeneous stresses,
  which might be the reason for these problems.  Treating the dimensionless
  shape parameter $\Psi$ within Mooney-Rivlin elasticity as an independent fit
  variable results in $\Psi \to 1$ for PVMAA/PVP, OTS, polyacrylamide CTAB,
  and polyacrylamide DTAB capsules.
  It is noticeable that all these materials
  give a Poisson ratio $\nu_\mathrm{2D} > 0.5$ employing the nonlinear Hookean
  fit. Only for the cucurbit[8]uril capsules, which have a Poisson ratio
  $\nu_\mathrm{2D} < 0.5$ we find $\Psi < 1$ and, thus, deviations from the
  neo-Hookean behavior. This indicates that most of the capsule materials
  discussed in this paper behave like a neo-Hookean material and, thus, also
  similar to a nonlinear Hookean material in the small strain limit. This limit is
  obviously applicable to PVMAA/PVP, OTS and polyacrylamide capsules, since
  they all satisfy $\vert \lambda_s - 1 \vert < 0.1$ over the whole contour,
  and also $\vert \lambda_\phi -1 \vert < 0.1$ in the non-wrinkled region. 
%  For
%  the wrinkled region we suggest that $\lambda_\phi$ should be allowed to show
%  stronger deviations because we actually use a different constitutive law in
%   this region, which is $\tau_\phi = 0$.  Due to these results, and also due
  Because of these results and 
  the fact that Mooney-Rivlin fits require a much higher computational cost,
    we focus on nonlinear Hookean elasticity
  in the following, which gives good results for all capsule types.

Where comparison to other rheological measurements is possible,
results from pendant droplet elastometry are in good agreement. 
For PMAA/PVP-capsules, the surface Young modulus of  $Y_\mathrm{2D} =
211\mathrm{mN/m}$ agrees with the findings in Ref.\ \cite{le2014interplay}.
In the following we will discuss results 
on the  OTS-, amino-functionalized polyacrylamide, and 
Span 65 capsules, which have not been previously discussed 
in the literature,  in more detail.
Pendant capsule elastometry allows us to obtain elastic moduli 
 of the two-dimensional 
capsule shell material for each  digitized image of the 
deflated  capsule shape (given at least one
  image of its undeformed reference shape). 
Therefore elastic moduli can be determined as a function 
of the deflation volume. 
If the volume change rate can be controlled,
elastic moduli can
 be determined as a function of the  volume  change
rate to investigate viscoelastic effects.
If series of images over one or several deflation cycles 
are available,  we  can investigate 
 aging effects, for example, 
by plastic deformation over many deflation cycles. 
 We will  explore these 
possibilities for  OTS-, polyacrylamide, and 
Span 65 capsules, starting with the latter.

%%%%%%%%%%%%%%%%%%%%%%%%%%
\subsection{Span 65 capsules}

\begin{figure*}[t!]
\centering
\includegraphics[width=\linewidth]{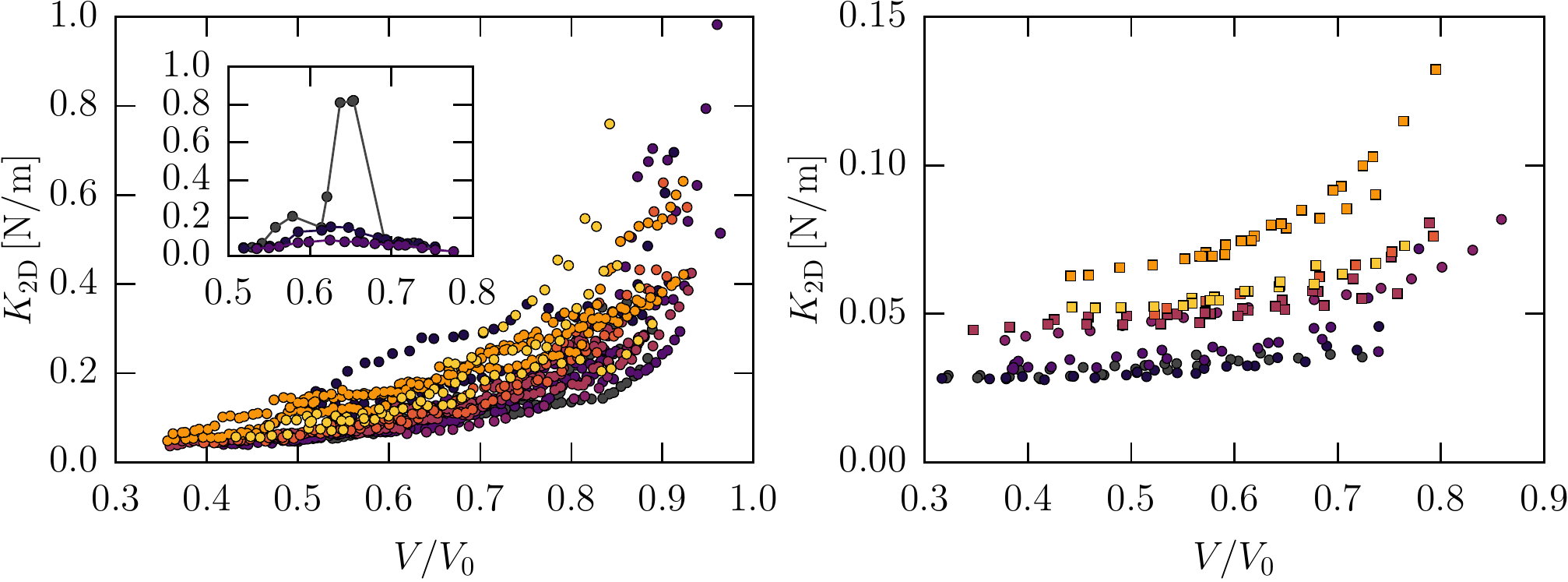}
\caption{
  \textbf{Left:} Pendant capsule elastometry 
  results for Span 65 capsules, i.e.,  
  $\mathrm{H}_2\mathrm{O}$-droplets coated with Span 65 (sorbitan
  tristearate) in dodecane. We analyzed eight individual
  capsules (color code), most of them fitted over three phases of deflation. 
  The inset also shows the three inflation phases for one of these capsules.
  The inflation phases differ significantly among the capsules,
  and there is no generic behavior as for the deflation phases.
  Among the individual capsules, volume change rates vary between $0.5$
  and $1.0\mu\mathrm{l/s}$.
  Surface shear-rheometry measurements give $K_\mathrm{2D} = 0.9\mathrm{N/m}$,
  Langmuir-Blodgett trough $K_\mathrm{2D} = 0.33\mathrm{N/m}$, spinning-drop
  $K_\mathrm{2D} = 0.36\mathrm{N/m}$, and rheoscope
  $K_\mathrm{2D}=0.54\mathrm{N/m}$. Rheological measurements are thus
  consistent with our method, which also reveals a strong variation of 
  $K_\mathrm{2D}$ with the volume. 
  The Poisson ratio is roughly given by
  $\nu_\mathrm{2D}=0.8$. 
  \textbf{Right:} 
  Pendant capsule elastometry results 
  for  polyacrylamide capsules, i.e., $\mathrm{H}_2\mathrm{0}$-droplet with
  $\mathrm{Na}_2\mathrm{CO}_3$,
  $\rm{N}\text{-}(3\text{-}\rm{Aminopropyl})\text{-}\rm{methacrylamide}$ and
  DTAB (circles) or CTAB (quads) surfactants. The outer phase 
   consists of \emph{p}-xylol and sebacoyl dichloride.  
  We analyzed four individual CTAB and four individual DTAB capsules, 
  most of them
  fitted over three phases of deflation.
  Deformations were applied after $60$ minutes
  equilibration time with the crosslinker. The Poisson ratio is
  $\nu_\mathrm{2D}=0.6$ with DTAB surfactants and $\nu_\mathrm{2D}=0.5$ with CTAB
  surfactants. The values of the area compression modulus are consistent with
  shear-rheometer measurements, which give $K_\mathrm{2D}=30 \dots
  100\mathrm{mN}/\mathrm{m}$.
}
\label{fig:span}
\end{figure*}

Span 65 has a polar head group connected to three carbon chains leading to
intermolecular interactions when adsorbed to a liquid interface.  Though not
explicitly crosslinked, the material shows elastic properties due to the
formation of temporary networks. For Span 65, our method agrees with four
different rheological measurements (surface shear-rheometry,
Langmuir-Blodgett, spinning-drop method, and shear flow rheoscope), which all
give area compression moduli $K_\mathrm{2D}$ between $0.3$ and $0.9$ N/m.
These fit well to the values $K_\mathrm{2D}=0.2\dots1.0\,\mathrm{N/m}$
obtained by our method for small deformations at $V/V_0 > 0.8$, see
Fig.\ \ref{fig:span} (left).  The pendant capsule elastometry results in
Fig.\ \ref{fig:span} (left) also reveal that the area compression modulus
strongly varies with the volume: deflated capsules with $V/V_0<0.5$ become
very soft with $K_\mathrm{2D}< 0.1\mathrm{N/m}$.  This pronounced compression
softening can eventually explain the deviations among previous rheological
measurements.
Upon re-inflating the capsule, the compression modulus
exhibits a non-monotonous behavior (see Fig.\ \ref{fig:span} inset)
but we do not find a generic pathway among the eight individual
capsules that we analyzed.
We can, however, speculate based on visual impressions from the images
that the capsule material develops overlaps or similar microscopic folds
that vanish after  complete re-inflation.
As a  consequence, we see hysteresis but no aging effects as the 
compression modulus returns to its original value 
after completing a deformation cycle, see  Fig.\ \ref{fig:span} (left). 
We also do not see a pronounced change of this behavior if the 
volume change rate is changed. 
All these results suggest that 
the compression softening could be a result 
of reversible rearrangements of the temporary network of the 
capsule material on time scales, 
which are short compared to the time scale of volume changes. 
These reversible rearrangements lead to an apparent 
decrease of elastic moduli with decreasing volume.
In Fig.\ \ref{fig:lambdas} we showed that Span 65 capsules
exhibit strongly inhomogeneous strains, which likewise
indicates a quite complex elastic behavior. Eventually,
one might conclude that Span 65 is not well described by
classical elastic models. Specific models that account for
the microscopic details of the material have to be developed
to analyze the elastic properties of Span 65 in more detail.
In contrast to permanently crosslinked polymer membranes 
Span 65 forms temporarily crosslinked network 
structures \cite{rehage2001ultrathin}.
Within these structures, the applied stresses can relax with time constants 
of the order of several minutes, which leads to time-dependent transitions
from solid into liquid like membranes. 
This more complicated rheological behavior, 
can only be described by time-dependent nonlinear constitutive laws.
Bending moduli can not be determined directly
from the images since the wrinkles are not visible by eye
although the shape analysis suggests the existence of wrinkles
over an extended region, see Fig.\ \ref{fig:figs} D. 
One could assume a wrinkle wavelength just 
below the image resolution, which gives
$\Lambda \leq 8 \,\mu \mathrm{m}$, $E_B\leq 2\cdot 10^{-20}\,\mathrm{Nm}$, and 
 $H\leq1.67\,\mathrm{nm}$.
Eventually wrinkles could also be absent in this system because compressive 
hoop stresses can be relaxed by the rearrangements in the  temporary network.

\begin{figure*}[t!]
\centering
\includegraphics[width=\linewidth]{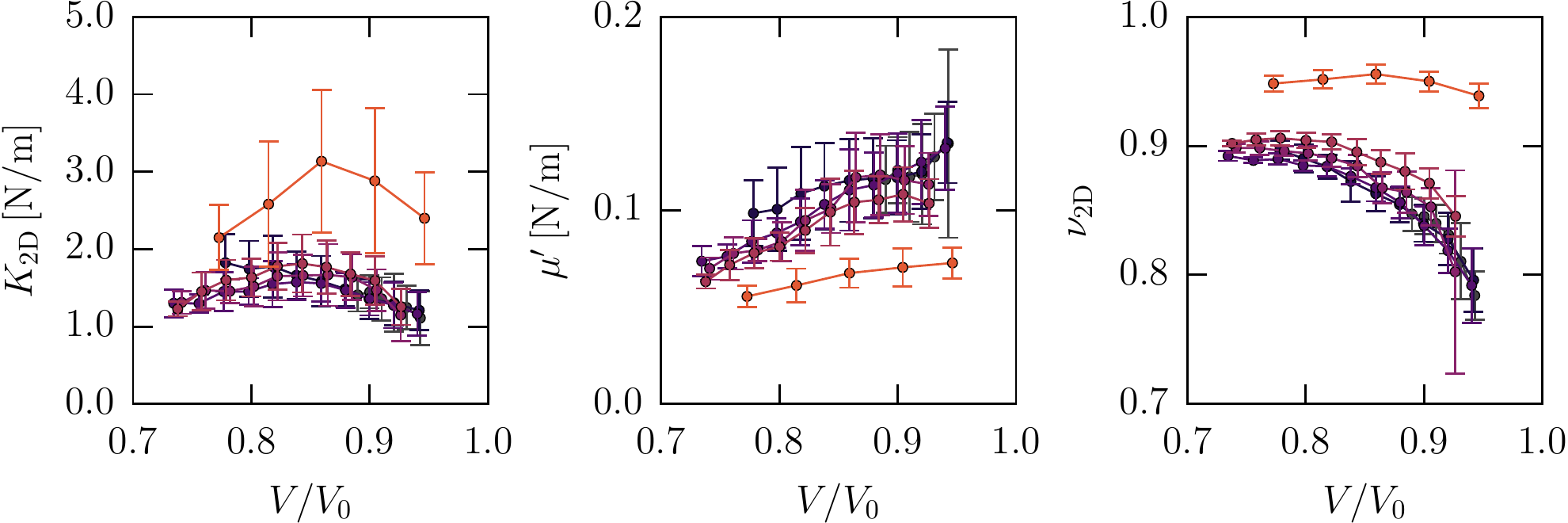}
\caption{
  Pendant capsule elastometry results for OTS-capsules, i.e., a 
  \emph{p}-xylol droplet in solution with 1,2,4-trichlorobenzene and coated
  with OTS in a glycerol-water
  mixture. We analyzed a single individual capsule with a 
  volume change rate of $0.5\,\mu\mathrm{l/s}$ (points) 
  and total volume reductions of 
  $2.5\,\mu\mathrm{l}$ (gray), $5.0\mu\mathrm{l}$ (blue), 
  $7.5\,\mu\mathrm{l}$ (dark violet),
  $10.0\,\mu\mathrm{l}$ (light violet), $12.5\,\mu\mathrm{l}$ (red).
  The same capsule was analyzed with step-wise increased
  volume change rates ($1.0$, $2.5$, $5.0$, $7.5\,\mu\mathrm{l/s}$) 
  up to $10\,\mu\mathrm{l/s}$ (orange)
  and the same total volume reductions. The capsule was
  subject to $30$ cycles of de- and inflation, and we analyzed
  $1674$ images for the first and last five cycles, 
  which allows us to calculate error bars by averaging over small
   volume ranges.
  Spinning drop measurements give $K_\mathrm{2D} = 3.0\dots
  7.5\,\mathrm{N/m}$ and rheoscope measurements $K_\mathrm{2D}=4.0\dots
  10\,\mathrm{N/m}$, depending on which Poisson ratio $\nu_\mathrm{2D}$ 
  is assumed to obtain $K_\mathrm{2D}$ 
  from the actually measured $Y_\mathrm{2D}$. 
  These values are slightly higher than our pendant elastometry measurements. 
  For the Poisson ratio we get roughly
  $\nu_\mathrm{2D}=0.85$, which is slightly above previous measurements
  predicting $\nu_\mathrm{2D}=0.5\dots 0.8$ \cite{koleva2012deformation}. 
  We see that, for the last $5$ cycles with a volume change rate 
  of $10.0\,\mu\mathrm{l/s}$ (orange), the material has 
  softened significantly, regarding $Y_\mathrm{2D}$ and $\mu^\prime$.
  The area compression modulus $K_\mathrm{2D}$ increased, however, 
  due to an increased Poisson ratio.
  In principle, these effects could either be induced by
  aging or by viscoelastic effects.
  For viscoelastic materials we typically expect a stiffening when
  volume change rates are increased. Therefore, we suggest that this softening
  is induced by aging at the intermediate
  rates $1.0$, $2.5$, $5.0$ and $7.5\,\mu\mathrm{l/s}$
54  that have been applied before the final $10.0\,\mu\mathrm{l/s}$ rate.
}
\label{fig:ots}
\end{figure*}

%%%%%%%%%%%%%%%%%%%%%%%%5
\subsection{OTS-capsules}

For the  OTS-capsules from Fig.\ \ref{fig:ots} 
we find values $K_\mathrm{2D}=1.0 \dots 4.0\,
\mathrm{N/m}$, which is just slightly below the rheological data 
from other methods giving
$K_\mathrm{2D}=3.0\dots10.0\,\mathrm{N/m}$ 
(spinning drop measurements give $K_\mathrm{2D} = 3.0\dots
  7.5\,\mathrm{N/m}$ and rheoscope measurements $K_\mathrm{2D}=4.0\dots
  10\,\mathrm{N/m}$), see Fig.\ \ref{fig:ots}.
In Fig.\ \ref{fig:ots}, we 
 analyzed a single OTS-capsule for different volume change rates
ranging from 
$0.5$ (slow) to $10.0\,\mu \mathrm{l}/\mathrm{s}$ (fast). 
In principle, this enables us to see viscoelastic effects. 
We expect a viscoelastic material to exhibit a smaller
shear modulus $\mu'$ for slow deformation such that creep 
or viscoelastic relaxation is possible. 
Fig.\ \ref{fig:ots} shows that 
the surface Young modulus and the shear modulus
$\mu'=K_\mathrm{2D}(1-\nu_\mathrm{2D})/(1+\nu_\mathrm{2D})$ 
are both significantly decreased for {\it higher}
volume change rates. 
Therefore, this is  probably an effect of aging
rather than viscoelastic behavior.
It is thereby difficult to explain the increased
area compression modulus and Poisson's ratio, which indicates
that the capsule material tends to become incompressible
due to microscopic effects, that cannot be observed in detail experimentally.
However, by exploring volume cycles for a wide range of volume change rates, 
it should, in principle, be possible to determine the frequency dependence
of the surface shear (storage) modulus $\mu'$ from these
measurements. Therefore, individual capsules should be prepared 
for each volume change rate to eliminate the influence of aging.

\begin{figure*}[t]
\centering
\includegraphics[width=\linewidth]{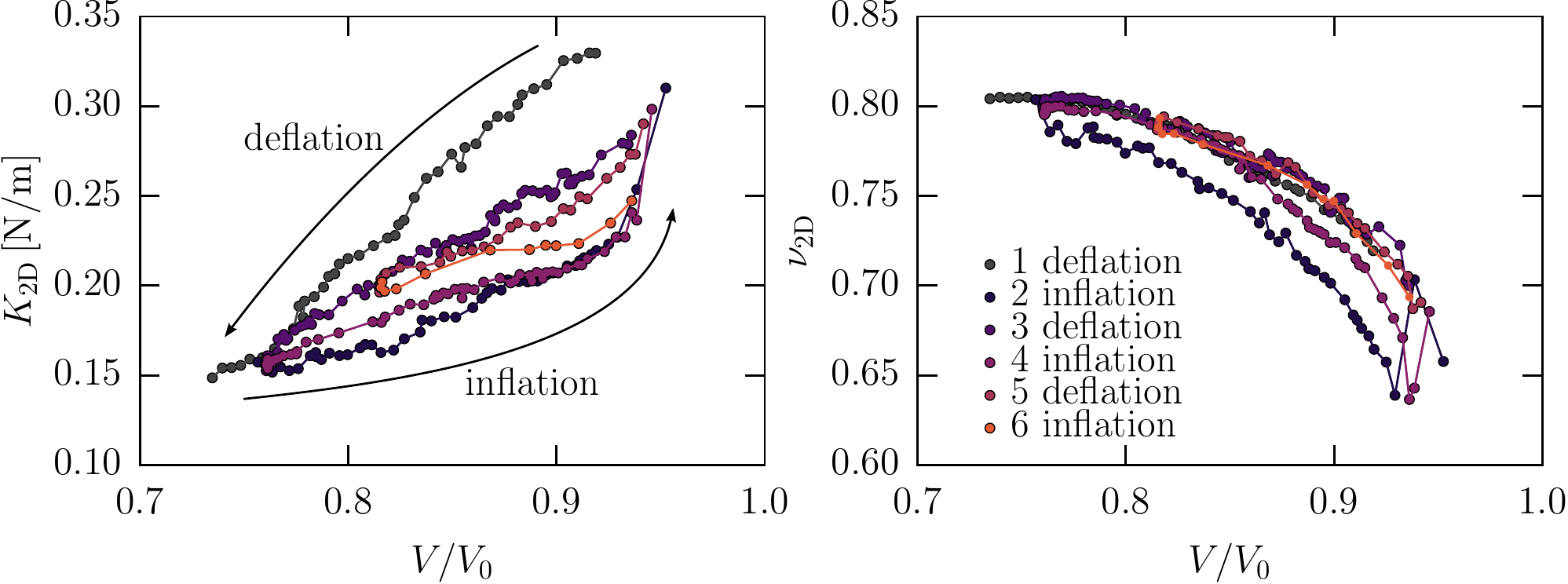}
\caption{
  We analyze a single OTS-capsule \cite{knoche2013elastometry}, i.e.,
  a $\mathrm{H}_2\mathrm{O}$-droplet coated with OTS in \emph{p}-xylol,
  for three consecutive deflation and inflation cycles. 
  This reveals aging effects: deformations become 
  nearly reversible after two complete deformation cycles. 
  For the first two cycles we clearly see hysteresis.
  Note that these OTS-capsules exhibit much smaller values of
  the area compression modulus $K_\mathrm{2D}$ compared to
  those in Fig.\ \ref{fig:ots}. Since both capsule membranes are made from
  OTS, and both should have a similar thickness ranging from
  $100$ to $1000\,\mathrm{nm}$, we conclude that this has to originate
  in the different liquid phases that have been used. Here, a water
  droplet was used in a \emph{p}-xylol phase, whereas for the capsules
  in Fig.\ \ref{fig:ots} a \emph{p}-xylol droplet was used in a gylcerin-water
  mixture. This possibly influences the network forming process, 
  such that capsules
  from Fig.\ \ref{fig:ots} appear more like an incompressible material.
}
\label{fig:multi}
\end{figure*}

For OTS-capsules (with different liquid phases compared to those in
Fig.\ \ref{fig:ots}) we analyzed
aging effects in more detail in Fig.\ \ref{fig:multi}
by monitoring the change of elastic constants over three consecutive 
deformation cycles of the same capsule. 
For this capsule, the deformation behavior becomes approximately 
reversible only after completing two deflation-inflation cycles. 
The first two cycles exhibit hysteresis hinting to 
plastic deformation in the capsule. Similar effects can 
be seen in Fig.\ \ref{fig:ots}, where the OTS-capsule
was subject to $30$ de- and inflation cycles in total.
The orange circles show the last five cycles where we
observe a softening regarding the shear modulus, a stiffening
regarding the area compression modulus, as well as an increased Poisson ratio.
For all quantities the volume dependence is weakened.
Capsules from Fig.\ \ref{fig:ots} obviously exhibit a more complex
aging behavior, which can only be caused by the different liquid phases
(essentially oil and water phase swapped),
because the OTS shell material is the same. 
However, these results suggests that by iteratively applying 
small volume change rates and small total volume reductions,
capsule deformations can reach a reversible regime, where
subsequent deformation cycles yield the same elastic constants
and aging is effectively absent. If, however, the volume change
rates or total volume reductions are successively increased,
aging proceeds and the capsule material becomes even softer.

By applying the wrinkling analysis 
to the capsules from Fig.\ \ref{fig:ots} we measure wrinkling
wavelengths $\Lambda = 0.2\,\mathrm{mm}$ leading to a 
bending modulus of $E_B = 10^{-15}\dots 10^{-14}\,\mathrm{Nm}$ 
and a thickness $H = 90\dots 290\,\mathrm{nm}$. 
Electron microscopy 
measurements give $H=100\dots1000\,\mathrm{nm}$, which roughly agrees. The
quality of these estimates depends crucially on the measurement of the
wrinkling wavelength since we have $E_B\propto \Lambda^4$. We prefer to measure
$\Lambda$ at the bottom of the wrinkles, because length measurements in the
center of the capsule can be better translated to the length scale prescribed
by the outer capillary diameter.

%%%%%%%%%%%%%%%%%%%%%%%%%%
\subsection{Polyacrylamide capsules}

We tested our software also 
on amino functionalized polyacrylamide capsules formed
with two different surfactants CTAB and DTAB, see Fig.\ \ref{fig:span} (right).
For the CTAB polyacrylamide capsules we find area 
compression moduli in the range
$K_\mathrm{2D}=50\dots 100\,\mathrm{mN/m}$ and a Poisson ratio
$\nu_\mathrm{2D}=0.5$, for the DTAB polyacrylamide capsules 
we find $K_\mathrm{2D}=30\dots
60\,\mathrm{mN/m}$ and $\nu_\mathrm{2D}=0.6$. These values are
consistent with surface shear-rheometry 
measurements giving $K_\mathrm{2D}=30\dots
100\,\mathrm{mN/m}$. Our pendant capsule elastometry results 
  show that the elastic properties of this type of capsule can
be tuned by changing only the surfactants (CTAB vs.\ DTAB) 
  and not the crosslinker. Fig.\ \ref{fig:span} (right) clearly shows 
 that CTAB gives consistently stiffer 
  capsules than DTAB. 
 Bending moduli for the DTAB polyacrylamide capsules are slightly smaller than
  for the CTAB polyacrylamide capsules.
 We find $E_B=10^{-14}\,\mathrm{Nm}$ for CTAB 
 and $E_B=8\cdot 10^{-15}\,\mathrm{Nm}$
  for DTAB (using images from Fig.\ \ref{fig:figs}).

%%%%%%%%%%%%%%%%%%%%%%
\section{Gibbs modulus}
\label{sec:Gibbs}

Finally, we like to discuss why the Gibbs modulus is not suitable
to analyze solid-like elastic shell materials.
As mentioned above, the Gibbs modulus 
$E_\mathrm{Gibbs} = \mathrm{d}\gamma/\mathrm{d}\ln A$  is frequently
determined from measurements of an ``effective surface tension'' $\gamma$ 
describing the solid shell interface as a function of the
surface area $A$. 
We use the term ``effective'' here, because, in principle, one cannot
determine a surface tension from deformations of a solid shell interface.
The Gibbs modulus  $E_\mathrm{Gibbs}$ is equal
to the area compression modulus $K_\mathrm{2D}$ for a fluid interface 
or for a two-dimensional solid interface
in a planar Langmuir-Blodgett trough geometry.

It is possible to
determine the Gibbs modulus by using a standard pendant drop tensiometer
measuring only the surface tension and the surface area. 
Commercial pendant drop tensiometers apply sine-like volume
changes and determine the complex Gibbs modulus with an elastic and a
viscoelastic contribution. This type of analysis 
is sometimes referred to as the ``oscillating drop'' method.

To be consistent with our purely elastic model, we
apply a linear fit in the $(\ln A,\gamma)$-plane. For the OTS-capsules we find
$E_\mathrm{Gibbs}=40\,\mathrm{mN/m}$, which is two
 orders of magnitude below the
actually measured area compression moduli. Similar results were obtained in
Ref. \cite{knoche2013elastometry}, where $E_\mathrm{Gibbs}$ and $K_\mathrm{2D}$
were compared for theoretically generated shapes. 
For the Span 65 capsules we get
$E_\mathrm{Gibbs}=33\,\mathrm{mN/m}$, 
which is one order of magnitude below the
value obtained in the Langmuir-Blodgett trough. 
We obtained similar values $E_\mathrm{Gibbs} \approx 40\,\mathrm{mN/m}$ 
by applying the oscillating drop method
to Span 65 capsules, which explicitly demonstrates that 
determining the Gibbs modulus
with a pendant drop tensiometer leads to misleading results,
as already stated in Ref.\ \cite{knoche2013elastometry}. 
The deformed shapes of elastic capsules cannot be fitted accurately with
the Laplace-Young equation for fluid interfaces as we have already shown in
Fig.~\ref{fig:network-formation}. 
Moreover, the relation between Gibbs modulus and 
 area compression modulus $K_\mathrm{2D}$  becomes non-trivial and 
geometry-dependent  because of inhomogeneous elastic stresses in the 
capsule geometry and  
the existence of a curved undeformed reference shape of the capsule.
Only in a planar geometry, 
where the elastically relaxed reference state is planar
 and where stresses remain homogeneous, the Gibbs modulus 
coincides with the area compression modulus.

%%%%%%%%%%%%%%%%%%%
\section{Conclusion}

We developed an efficient and completely automated  C/C++ software
in order to perform pendant capsule elastometry in 
 pendant drop devices. 
The analysis is based on a thin elastic shell model of the 
capsule interface and, thus, applies  for elastic solid
capsule materials. Such materials can 
 be recognized, for example,  by their ability to develop wrinkles.

The analysis requires a reference shape, for which we usually
assume that it can be described by a liquid Laplace-Young shape
if the shell material is crosslinked at a liquid interface.  
A minimum set of two
images, one of the reference shape and another of a 
 deformed shape, and two experimental parameters, namely the
density difference $\Delta\rho$ and the outer capillary diameter $b$, are
sufficient to run the complete analysis and obtain values for 
Young's modulus and Poisson's ratio using the Hookean elasticity model, 
or, alternatively, the
Rivlin modulus and the dimensionless shape parameter using the Mooney-Rivlin
elasticity model. In addition, if wrinkling occurs, 
the bending modulus and, thus, the shell thickness
can be determined from a wrinkle wavelength measurement.

 More interesting results are often
obtained, if a whole sequence of deformed states can be analyzed 
 in chronological
order, which makes it possible to investigate the dependence of the elastic
moduli on the capsule volume.  This is where possible phase
transitions, hysteresis and aging effects, or plastic deformations
of the material could  be detected.
Future work should explore whether a frequency-dependent 
surface shear (storage) modulus $\mu'$ can be reliably 
determined by measurements 
at different volume change rates. 

In our present implementation 
pendant capsule elastometry does not require  experimental information 
on the pressure $p$ but rather serves itself as a 
pressure measurement because 
$p$ is determined by shape fitting. 
If such pressure is measured independently, 
this additional information can be readily used 
to improve the results for the remaining fit parameters
by setting $p$ to a constant value within shape regression.
This option is supported by the current version of our software.
But there are also other possibilities to use the pressure information,
e.g., to directly calculate local stresses in the deformed state 
from additional image information 
on local curvatures and radii \cite{danov2015capillary}.
This additional stress information can then  be used to improve 
the shape fitting procedure beyond simple elimination 
of one fit parameter \cite{Nagel2017}.

In the present form of the code, we provide linear Hookean elasticity,
  nonlinear Hookean elasticity, and Mooney-Rivlin or neo-Hookean elasticity to
  describe the elastic behavior of the shell material. We find that the
  nonlinear Hookean model gives the best compromise between accuracy and
  performance.  Moreover, we randomly checked the results against
  Mooney-Rivlin elasticity (which is much slower due to numerical
  determination of the closure relations).  This revealed that both elasticity
  models give similar results over a wide range of materials and deformation
  behavior.  
  We find that the simple linear Hookean elasticity can only
  describe very small deformations compared to nonlinear Hookean or
  Mooney-Rivlin elasticity. We gained this insight from generating theoretical
  shapes, where we decreased the pressure successively, thereby simulating
  proceeding deflation. Numerics for the linear Hookean elasticity failed much
  earlier in fulfilling the required boundary conditions, which is necessary
  to generate a valid shape.  It remains to be verified
  systematically, however, 
   what differences eventually arise between fitting with
  different elasticity models.
% This is justified as long as strains remain small
% but the method can be easily adapted to other elastic models, such as 
% Mooney-Rivlin elasticity for polymeric materials. 
%\Cr{In a recent work \cite{Nagel2017}, Hencky strain measures are 
%proposed.}
% Such models can be more appropriate for materials exhibiting stretch 
% or compression softening as typically observed for 
% many capsule materials, see Fig.\ \ref{fig:span} 
% for Span 65 and polyacrylamide capsules. 
Fitting with the most appropriate 
model should produce the least elastic 
parameter variation as a function of the volume.

For certain materials 
the use of even more specific elastic models is more appropriate,
for example, hydrophobins coating air bubbles
\cite{Aumaitre2013,knoche2013elastometry} act as interfacial 
rafts of hard particles with soft shells, which require a more 
refined elastic description to interpret capsule shapes 
correctly and avoid jumps in elastic constant in elastometry 
fits \cite{Knoche2015}.
Therefore, future work should  also aim at implementing 
different elastic models in order to compare fit results 
for different  models and  determine the most appropriate model from 
the data. In particular, for the Span 65 investigated in this paper,
we suggest to develop such microscopic models, because we found
strongly inhomogeneous strains and generally atypical elastic behavior.
Moreover fit results revealed a pronounced compression softening, which 
we did not find for any other capsule material, which also hints to the
use of a more appropriate elasticity model.

As we have shown, our software for pendant drop devices is widely applicable.
We tested it on different materials and the results are in good agreement
with available  rheological data.
We make the  OpenCapsule software freely 
 available under the GPL license  \cite{gplv3} at 
{\tt github.com/jhegemann/opencapsule}.
A user manual is also available as Supplementary material.

%%%%%%%%%%%%%%%%%%%
\section{Acknowledgements} 

We thank Sandrine Le Tirilly and C\'ecile Monteux 
%(\'Ecole Sup\'erieure de
%Physique et de Chimie Industrielles de la Ville de Paris (ESPCI)) 
for
providing capsule images from capsules used in Ref.\ \cite{le2014interplay}
(Fig.\ \ref{fig:figs} B) and Andrew Salmon and Chris Abell 
%(Department of Chemistry, University of Cambridge) 
for providing images from capsules used in
Ref.\ \cite{Salmon2016} (Fig.\ \ref{fig:figs} A).  
Moreover we thank Patrick Degen 
%(Department of Chemistry, TU Dortmund University)
for providing the capsule images analyzed in Figs.\ \ref{fig:multi}. 
We thank Horst-Holger Boltz 
%(Department of Physics, Georg-August-University G{\"o}ttingen) 
and Tobias Kampmann 
%(Department of Physics, TU Dortmund University) 
for a careful reading of the manuscript and fruitful discussions.

%%%%%%%%%%%%%%%%%%
\appendix

%%%%%%%%%%%%%%%%%%%%%%
\section{Least squares}
\label{sec:leastsquares}

Both the shooting method and the shape regression require 
least square error/deviation minimization. Fitting shape 
equations to experimental contours is implemented by a
nested minimization algorithm, which minimizes the distance between 
individual shape segments (and the boundary conditions) in 
an inner loop, which we call \textit{shooting method} (see Sec.\ 
\ref{sec:shooting} below), 
and the deviation between theoretical 
shape and contour in an outer loop, which we 
call \textit{shape regression}   (see Sec.\  \ref{sec:regression} below).
At first, we characterize the error function that has to
be minimized within the least square algorithm.

Consider a global residual vector 
\begin{eqnarray}
\vec{F}(\vec{x}) &=& (\vec{\varphi}_1(\vec{x}),\dots,\vec{\varphi}_N(\vec{x}))^T
\end{eqnarray}
assembling the individual residuals
$\vec{\varphi}_i$ that depend on an arbitrary parameter set $\vec{x}$. 
In case of the shooting method, the residuals $\vec{\varphi}_i$ are 
defined by the distances between consecutive shape segments and, finally,
between the last shape segment and the boundary condition. Thus, the
parameter set $\vec{x}$ can be identified with the set of shooting parameters,
which are the initial values of the individual shape segments.
In case of the shape regression, the residuals $\vec{\varphi}_i$ 
give the shortest
distances between the discrete points of the contour (obtained from the image) 
and the theoretical shape profile given by a solution of the shape equations.
Thus, the parameter set $\vec{x}$ can be identified with the shape parameters
characterizing the solutions of either the 
Laplace-Young equations \eqref{eqn:laplace-young}
or the elastic shape equations \eqref{eq:shape}.

 We now introduce the general least square method,  stressing
  again that we use this method for both the shooting method \textit{and} the
  shape regression.  The Jacobian $\op{J}_{\vec{F}}$ measures the change of
  $\vec{F}(\vec{x})$ at some point $\vec{x}$.  In order to minimize the
  euclidean norm $\Vert \vec{F}(\vec{x})\Vert$ with respect to $\vec{x}$ we
  linearize $\vec{F}(\vec{x})$ within a small region $\Delta \vec{x}$
  according to
\begin{eqnarray}
\Vert \vec{F}(\vec{x}+\Delta\vec{x}) \Vert &=& 
  \Vert \vec{F}(\vec{x}) + \op{J}_{\vec{F}}\Delta\vec{x} \Vert 
   \hspace*{2mm}\stackrel{!}{=}\hspace*{2mm}0
\end{eqnarray}
yielding the linear and typically over-determined system of equations
\begin{eqnarray}
\op{J}_{\vec{F}}\Delta\vec{x} &=& -\vec{F}(\vec{x}).
\label{eqn:shift}
\end{eqnarray}
Standard algorithms like the Gauss-Newton method solve the quadratic normal 
equation 
\begin{eqnarray}
\op{J}_{\vec{F}}^T\op{J}_{\vec{F}}\Delta\vec{x} &=& -\op{J}_{\vec{F}}^T\vec{F}(x),
\end{eqnarray}
but we prefer to directly solve \eqref{eqn:shift}, because the 
condition of $\op{J}_{\vec{F}}^T\op{J}_{\vec{F}}$ can be poor in comparison to the
condition of $\op{J}_{\vec{F}}$, i.e.,
\begin{eqnarray}
\mathrm{cond}(\op{J}_{\vec{F}}^T\op{J}_{\vec{F}}) &\sim& 
  \mathrm{cond}(\op{J}_{\vec{F}})^2.
\end{eqnarray}
We do so by decomposing 
$\op{J}_{\vec{F}} = \op{Q}\op{R}$ via Householder transformations
and multiplying with $\op{Q}^T$,
\begin{eqnarray}
\op{R}\Delta\vec{x}&=&-\op{Q}^T\vec{F}(x),
\end{eqnarray}
where we used that $\op{Q}^T\op{Q}=\mathbbm{1}$.
Note that this yields
\begin{eqnarray}
\op{R} &=& \binom{\op{R_0}}{0} 
\end{eqnarray}
and 
\begin{eqnarray}
\op{Q}^T \vec{F}(x) &=& \binom{\vec{b}_0}{\vec{b}_1}
\end{eqnarray}
in case of over-determined systems, such that
the solution is given by
\begin{eqnarray}
\Delta\vec{x} &=& \op{R}_0^{-1} \vec{b}_0
\label{eqn:final-shift}
\end{eqnarray}
with a finite error 
\begin{eqnarray}
\Vert \vec{F}(\vec{x}) + \op{J}_{\vec{F}}\op{R}_0^{-1} \vec{b}_0 \Vert 
    &=& \Vert \vec{b}_1 \Vert.
\end{eqnarray}

Applying the parameter shift $\Delta\vec{x}$
resulting from \eqref{eqn:final-shift} or \eqref{eqn:shift} 
iteratively to the current parameter set
finally gives a solution $\vec{x}^*$, which minimizes 
$\Vert \vec{F}(\vec{x}) \Vert$, i.e., 
\begin{eqnarray}
\min_{\vec{x}} \Vert \vec{F}(\vec{x})\Vert &=& \vec{x}^*.
\end{eqnarray}
To obtain the sequence $\{\vec{x}^k\}$ that finally converges to $\vec{x}^*$, 
we use the update scheme
\begin{eqnarray}
\vec{x}^{k+1} &=& \vec{x}^k + \lambda_j \Delta \vec{x}^k,
\end{eqnarray}
where $\lambda_j$ is chosen such that $\Vert \vec{F}(\vec{x})\Vert$ decreases
in each step of iteration.  Several line search methods may be applied here,
but, in view of efficiency, we prefer to chose $\lambda_j = 1/2^j$, where we
increase $j$ starting from $j=0$ until
\begin{eqnarray}
\Vert \vec{F}(\vec{x}^k+\lambda_j\Delta\vec{x})\Vert 
  &<& \Vert \vec{F}(\vec{x}^k)\Vert.
\end{eqnarray}
This is sometimes referred to as a ``backtracking line search'' method.
The minimization algorithm will be used for the multiple shooting method,
where \eqref{eqn:shift} is quadratic, as well as for the final regression of
the shape equations, where \eqref{eqn:shift} is strongly over-determined.
Since we exclusively use numerical differential quotients the algorithm
converges linearly, whereas a classical Newton minimization would converge
quadratically due to analytical derivatives.

%%%%%%%%%%%%%%%%%%%%%%%%%
\section{{Reference shape}}
\label{sec:app-reference}

Obtaining a shape profile from the set of shape equations
\eqref{eqn:laplace-young} is trivial, since there are no shooting
parameters. In practice, one integrates the set of shape equations
\eqref{eqn:laplace-young} (while increasing the arc length $s_0$) until
$r_0(s_0)= a/2$ is satisfied for the second time, meaning that the shape
enters the capillary from $r_0 > a/2$ (there is also a solution, 
which enters the capillary for the first time from $r_0<a/2$; this solution 
has a much smaller volume and does usually not correspond to the 
experimental reference shape). 
The arc length $s_0$ that satisfies
this condition is chosen as the undeformed contour length $L_0$, such that
$r_0(L_0) = a/2$. Note that the undeformed length $L_0$ is fixed for the
deformed shape profiles.  The resulting reference shape
\begin{eqnarray}
\vec{y}_0(s_0, \vec{x}_0) &=& \begin{pmatrix} r_0(s_0, \vec{x}_0)\\
       z_0(s_0, \vec{x}_0) \\ \psi_0(s_0, \vec{x}_0) \end{pmatrix}
\end{eqnarray}
is obtained as a function of of the parameter set 
$\vec{x}_0 = \{p_0, \rho, \alpha\}$, 
which are adapted during the shape regression 
(see Sec.\ \ref{sec:regression}) 
to optimally match the contour points extracted from the image.

\vspace*{2em}

%%%%%%%%%%%%%%%%%%%%%%%%%
\section{Shooting method}
\label{sec:shooting}

Solving the elastic shape equations requires a shooting method to be applied,
because of the unknown initial tension $\tau_s(0) = \mu$ at the capsule's apex.
For a given initial value $\mu$ we therefore integrate the shape
equations starting at the capsule's apex from $s_0=0$ to $s_0 = L_0$, where
$L_0$ was determined before by satisfying the boundary condition 
of the Laplace-Young reference shape. We thereby obtain a deformed 
shape trajectory $\vec{y}(s_0;\mu)^T$,
which depends on the reference shape via the shape profile 
$r_0(s_0)$ and the length
of the undeformed contour $L_0$. However, the
deformed length $L = \int_0^{L_0} \lambda_s \mathrm{d}s_0$ of 
course adapts according to the
stretch factor $\lambda_s$. 
For a capsule with inner capillary width $a$ centered at $r=0$ 
a valid solution has to satisfy the boundary condition
\begin{eqnarray}
f(\mu) &=& r(L_0; \mu) - a /2 \hspace*{2mm}\stackrel{!}{=}\hspace*{2mm} 0.
\label{eq:boundary-deviation}
\end{eqnarray}
The function $f(\mu)$, which is measured from the solution
 $\vec{y}(s_0;\mu)^T$,
has to be minimized by applying a bisection with respect to the parameter
$\mu$.  We recommend a bisection in this case, because the function $f(\mu)$
is very steep, particularly for large area compression moduli
$K_\mathrm{2D}$. The algorithm is assumed 
to be converged if $\vert f(\mu) \vert < \epsilon_\mathrm{single}$.
Note that our software takes this as a minimum criterion, i.e., it tries
to minimize $\vert f(\mu) \vert$ even further until the interval within
the bisection method becomes smaller than $10^{-16}$. It is generally important
to minimize $\vert f(\mu) \vert$ as far as possible, because the shape 
trajectories are very sensitive to the initial value $\mu$.

In cases where the required accuracy $\epsilon_\mathrm{single}$ can not be
reached, we further improve solutions by applying a multiple shooting method
subsequently. For this purpose we divide the interval $[0, L_0]$ in $q$
sub-intervals with $q+1$ grid points at $s_k = k\,L_0/q$, where
 $k=0,\dots,q$.  On the sub-interval $[s_k,\,s_{k+1}]$ 
we define the $k$-th segment
\begin{eqnarray}
\begin{split}
\vec{y}_k(s_0) &\equiv& \vec{y}_k^0+ \int_{s_k}^{s_0}\mathrm{d}s_0' 
   \,\vec{f}\left(\vec{y}(s_0') \, ; \, \vec{y}_k^0\right)
\end{split}
\end{eqnarray}
by integrating the set of shape equations 
\begin{eqnarray}
\vec{y}' &=& \vec{f}\left(\vec{y}(s_0) \, ; \, \vec{y}_k^0\right),
\end{eqnarray}
starting at
\begin{eqnarray}
\vec{y}_k^0 &=& (r_k^0, z_k^0, \psi_k^0,\tau_k^0) \in \mathbb{R}^4
\end{eqnarray}
and ending at
\begin{eqnarray}
\vec{y}_k &\equiv& \vec{y}_k(s_{k+1},y_k^0) \in \mathbb{R}^4.
\end{eqnarray}
Note that the final segment $\vec{y}_{q-1}$ has to match the 
final grid point $\vec{y}_{q}$.
Having decomposed the continuous solution in $q$ individual segments,
we track the $k$-th residual vector separating the
 $k$-th and $(k+1)$-th segment via
\begin{eqnarray}
\vec{\varphi}_k &=& 
\begin{cases}
\vec{y}_k-\vec{y}_{k+1}^0 \in \mathbb{R}^4  & k=1\dots q-2 \\
\vec{y}_{k,1}-\vec{y}_{k+1,1}^0 \in \mathbb{R}^1 & k=q-1\,.
\end{cases}
\end{eqnarray}
The last segment and grid point, $\vec{y}_{q-1,1} = r_{q-1}(s_{q})$ 
and $\vec{y}_{q,1}^0=r_q^0= a/2$,
define the final boundary condition \eqref{eq:boundary-deviation}, 
where the capillary
has to be matched. Any other residual corresponds to continuity 
conditions that ensure a smooth shape.
To arrange the segments into a continuous solution while satisfying 
the boundary condition at the
capillary, we have to set up the
Jacobians for each segment $\vec{y}_k$, where $k\in[0,q-1]$, 
with respect to the 
corresponding initial values $\vec{y}_k^0$.
At $s_0$ only $\tau_0^0$ can be chosen freely, 
whereas $r_0^0 = 0$, $z_0^0=\zeta$ and
$\psi_0^0=0$ are fixed due to axis symmetry. At $s_q$ we have to satisfy the
boundary condition $r_{q-1}(s_{q}) - a/2 = 0$ whereas
$z_{q-1}(s_{q})$, $\psi_{q-1}(s_{q})$ and $\tau_{q-1}(s_{q})$ are arbitrary.  
The Jacobian
$\op{J}_0$ corresponding to $\vec{y}_0$ is a column vector in $\mathbb{R}^4$,
the Jacobian $\op{J}_{q-1}$ corresponding to $\vec{y}_{q-1}$ is a row vector in
$\mathbb{R}^4$.  All intermediate Jacobians $\op{J}_k$ with $k=1,\dots,q-2$
are quadratic matrices in $\mathbb{R}^{4\times 4}$ and we can write them as
\begin{eqnarray}
 \op{J}_0 &=& \frac{\partial \vec{y}_0}{\partial \tau_0^0},
     \hspace*{1em} \op{J}_k 
 = \frac{\partial \vec{y}_k}{\partial \vec{y}_k^0}, 
  \hspace*{1em} \op{J}_{q-1} = 
   \frac{\partial r_{q-1}}{\partial \vec{y}_{q-1}^0}\,,
\end{eqnarray}
where we use differential quotients
\begin{eqnarray}
	\frac{\partial \vec{y}_{k}}{\partial \vec{y}_{k,i}^0} &=&
\frac{1}{2\Delta}\left(\vec{y}(s_{k+1};s_k,\vec{y}_{k}^0+\Delta\vec{e}_i) 
  \right. \\
 &-& \left. \vec{y}(s_{k+1};s_k,\vec{y}_{k}^0-\Delta\vec{e}_i) \right)
\nonumber
\end{eqnarray}
with canonical unit vectors $\vec{e}_i$ and $i=1,\dots,4$. 
Note that we typically use $\Delta = 10^{-6}$.
Finally we find the block-matrix 
\begin{align}
\begin{split}
 \op{J} &= \frac{\partial (\vec{y}_0-\vec{y}_1^0,\,\dots,
 \vec{y}_{q-1,1} -\vec{y}_{q,1}^0)}{\partial (\vec{y}_0^0,\,\dots, 
  \vec{y}_{q-1}^0)} \\
&=
\begin{pmatrix}
\op{J}_0  &  -\mathbbm{1} & \hdots & 0 \\
\vdots  & \ddots & \ddots &  \vdots\\
\vdots &  & \ddots & -\mathbbm{1} \\
0  & \hdots  & \hdots & J_{q-1} \\
\end{pmatrix}\,,
\end{split}
\end{align}
where $\mathbbm{1}\in\mathbb{R}^4$ denotes the identity matrix.
Applying the least square minimization method described above, 
i.e.,  solving the quadratic system $\op{J}_{\vec{F}}\Delta \vec{x} = -\vec{F}$
iteratively, where 
 \begin{eqnarray}
 \vec{F} = (\vec{\varphi}_0,\dots,\vec{\varphi}_{q-1})
 \end{eqnarray}
 assembles the residuals and 
\begin{eqnarray}
\Delta \vec{x}=(\Delta \tau_0^0, \Delta\vec{y}_1^0,\dots, \Delta\vec{y}_{q-1}^0)
\end{eqnarray}
is the initial value shift we get in each iteration, 
we finally converge into the continuous solution.
The speed of convergence varies with the number of sub-intervals $q$,
which thus has to be optimized in each iteration. 
We typically increase $q$ corresponding to $q\to q+4$ starting at $q=4$
until we achieve convergence. This is efficient, because it keeps $q$ small.
Note that adding only a single interval, i.e., $q\to q+1$, leads to four
extra dimensions in the quadratic system $\op{J}\Delta \vec{x} = -\vec{F}$.

The multiple shooting is assumed to be converged if $\Vert \vec{F} \Vert <
\epsilon_\mathrm{multi}$, which also implies $\vert f(\mu)\vert <
\epsilon_\mathrm{multi}$. It is thus reasonable to use
$\epsilon_\mathrm{single} = \epsilon_\mathrm{multi}$. Note that the multiple
shooting method has to be applied only if the required accuracy in the single
shooting method could not be reached.

The resulting deformed shape
\begin{eqnarray}
\vec{y}(s_0, \vec{x}) &=& \begin{pmatrix} r(s_0, \vec{x})\\
   z(s_0, \vec{x}) \\ \psi(s_0, \vec{x})
   \\ \tau_s(s_0,\vec{x}) \end{pmatrix}
\end{eqnarray}
is obtained as a function  of the parameter set 
$\vec{x} = \{p, \nu, K_\mathrm{2D}\}$, 
which are adapted during the shape regression 
(see Sec. \ref{sec:regression}) 
to optimally match the contour points extracted from the image.

%%%%%%%%%%%%%%%%%%%%%%
\section{Shape regression}
\label{sec:regression}

In  the shape regression we find 
the material parameters which minimize the deviation/error between 
contours and theoretical shapes from solving shape equations.

The Laplace-Young equation depends on the parameter set $\vec{x}_0 = (p_0,
\Delta\rho, \alpha)$, where $\alpha$ is a scaling factor, which sets the length
scale.  The elastic shape equations depend on the parameter set 
$\vec{x} = (p, \nu_\mathrm{2D},K_\mathrm{2D})$. Let
$(\hat{r}_i,\hat{z}_i)$ with $i=1,\dots ,N$ be a set of 
contour points resulting from image processing. We translate
this set of contour points, such that these are symmetric with
respect to the $z$-axis and the apex is located at $z=0$. We then
accordingly chose $z(0) = \zeta = 0$. We thereby fix the theoretical
shape relatively to the contour points at the apex, and minimize the 
residual along the remaining shape profile.

The residuals
\begin{eqnarray}
\vec{\varphi}_i &=& \min_{s_0\in [0, L_0]} 
\begin{pmatrix}
\vert \hat{r}_i \vert - r(s_0, \vec{x}) \\
\hat{z}_i - z(s_0, \vec{x})
\end{pmatrix}
\label{eq:phi-error}
\end{eqnarray}
are calculated by a bisection-like algorithm in the arc length $s_0$,
which terminates when the interval length falls below the threshold $\epsilon_\mathrm{rms}$.
From the residuals $\vec{\varphi}_i$ we calculate the average mean square deviation
\begin{eqnarray}
\chi &=& \sqrt{\frac{1}{N}\sum_{i=1}^N \Vert \vec{\varphi}_i\Vert^2}
\label{eq:error}
\end{eqnarray}
between the contour and the theoretical shape, as well as the Jacobians
\begin{eqnarray}
\mathsf{J}_0 &=& \frac{\partial(\vec{\varphi}_{1},
   \dots ,
   \vec{\varphi}_{N})}{\partial (p_0, \Delta\rho, \alpha)} \\
\mathsf{J} &=& \frac{\partial(\vec{\varphi}_{1},
   \dots ,
   \vec{\varphi}_{N})}{\partial (p_0, \nu_\mathrm{2D}, K_\mathrm{2D})}
\end{eqnarray}
for the reference and the deformed shape.  These Jacobians are sufficient to
minimize the error $\chi$ and find the best fit parameter set $\vec{x}$
by solving the strongly over-determined system 
$\mathsf{J}_{\vec{F}}\Delta\vec{x} = -\vec{F}$ iteratively.
Note that we have to find the best fit parameter set for the reference
shape first, and afterwards perform the shape regression for the
deformed shape, using the already determined reference shape.
Each iteration of shape regression requires three
numerical derivatives to find the elements of the Jacobian, which in turn
require two executions of the shooting method.
This yields a parameter shift $\Delta\vec{x}$
in each iteration, which we assume to be converged if 
we find $\lambda_j \Vert \Delta \vec{x} \Vert < \epsilon_\mathrm{laplace/hooke}$
during the backtracking line search.
In addition to the minimization
algorithm explained above, our software additionally provides 
the so-called Nelder-Mead downhill
simplex method, which works without derivatives. In rare cases, where the 
standard procedure fails, one should try this more robust method.

%%%%%%%%%%%%%%%%%%%%%%%%%%%%%
\section{Numerical thresholds}
\label{sec:thresholds}

To ensure convergence of the shape regression and the shooting method, 
we have to specify
thresholds. 

For the average mean square
displacement \eqref{eq:error}, i.e., the individual
residuals \eqref{eq:phi-error} between the contour points 
and the theoretical shape we apply a bisection-like
algorithm terminating when the interval length falls 
below the threshold $\epsilon_\mathrm{rms}$.

For the single and multiple shooting methods we define 
the thresholds $\epsilon_\mathrm{single}$
and $\epsilon_\mathrm{multi}$, which have different meanings: the accuracy 
$\epsilon_\mathrm{single}$ is reached if 
$\vert f \vert < \epsilon_\mathrm{single}$, 
see eq.\ \eqref{eq:boundary-deviation},
is satisfied for the boundary deviation at the capillary,
whereas the accuracy $\epsilon_\mathrm{multi}$ is reached if 
$\Vert\vec{F}\Vert < \epsilon_\mathrm{multi}$ is satisfied for the global
residual, 
which also implies $\vert f \vert < \epsilon_\mathrm{multi}$.
We define $\epsilon_\mathrm{laplace}$
and $\epsilon_\mathrm{hooke}$ as thresholds for the euclidean norm of 
the parameter shift $\lambda_j\Delta \vec{x}$, which is applied to the
parameters of the shape 
equations during the regression  and the
backtracking line search, respectively. 
To integrate the shape equations we use
a $4$-th order Runge-Kutta method with constant 
step widths $h_\mathrm{laplace}$ and $h_\mathrm{hooke}$.

In Tab.\ (\ref{tab:num}) 
 standard values for the numerical algorithms are given. 
For the analysis of the capsules used in this paper, the numerical thresholds
always ranged within the given boundaries.
To improve the performance
for specific capsules these thresholds can be increased,
 but it should be checked
if the results are still in rough agreement with higher 
precision measurements, meaning that
no systematic errors occur. Note that the parameters of the image 
processing also change the numerical behavior
since the set of contour points results directly from image processing. 
Changing, for example, the 
width of the Gaussian smoothing of 
the image will alter the fitting results.

\begin{table}
\centering
\begin{tabular}{p{0.3\linewidth}|p{4em}|p{4em}}
symbol & precision & performance \\
\hline
$\epsilon_\mathrm{rms}$ & $10^{-16}$ & $10^{-16}$ \\
$\epsilon_\mathrm{single}$ & $10^{-6}$ & $10^{-4}$\\
$\epsilon_\mathrm{multi}$ & $10^{-6}$ & $10^{-4}$ \\
$\epsilon_\mathrm{laplace}$ & $10^{-6}$ & $10^{-4}$ \\
$\epsilon_\mathrm{hooke}$ & $10^{-6}$ & $10^{-4}$ \\
$h_\mathrm{laplace}$ & $10^{-4}$ & $10^{-3}$ \\
$h_\mathrm{hooke}$ & $10^{-4}$ & $10^{-3}$
\label{tab:num}
\end{tabular}
\caption{Precision and performance optimized values
for the thresholds used in the numerical algorithms.}
\end{table}

%%%%%%%%%%%%%%%%%%%%%%%%%%%%%%%%%
\section{Image processing and requirements}
\label{sec:image}

Several filters, transformations and algorithms are applied to the image in
order to get a set of contour points, which can be used for shape regression.
Initially, we use a Gaussian filter to smoothen the image and run the Canny
edge detection. This is common practice to extract contours from images.  From
the binary image we measure the outer and inner capillary diameter
(the latter implicitly in terms of the fit parameter $\alpha$), which is
necessary to relate the length scale set in the image to SI units. Likewise, 
we measure the height of the capsule and its distance from the bottom of
the image. These quantities are necessary to translate the contour points
according to the remarks stated in \ref{sec:regression}.
Furthermore, we extract the contour points and reduce their number to
improve efficiency.  To ensure that the capturing algorithm works correctly,
images have to meet certain requirements.  In principle, all file formats
supported by the OpenCV library can be used with our software, but we
recommend png-files.  Gravity should act downwards along the vertical axis and
the capsule should be centered in the image with the capillary entering the
image at the top.  If these requirements are fulfilled, it is, in contrast to
the typical pendant drop software packages, not necessary to select the
capsule region manually, since the software detects the capillary and
therefore the top side of the capsule automatically. The background should be
uniformly colored and clean from small particles or other objects disturbing the
edge detection. To ensure a proper automatic wrinkle detection the wrinkles
should be visible over the whole width of the capsule.  If the edge detection
for the wrinkles does not work, one can provide a manually measured wrinkle
wavelength in the configuration file. Even if the edge detection for the
wrinkles works, one should randomly check the results by measuring the
wrinkling length manually since the automatic detection requires uniformly
illuminated capsules.

\vspace*{2em}

\bibliographystyle{elsarticle-num} 
\bibliography{lit}

\end{document}